\documentclass[manuscript]{aastex63}

\received{January 05, 2022}
\revised{February 11, 2022}
\accepted{XXXX YY, 2022}
\submitjournal{AJ}
\shorttitle{The open clusters NGC 1664 and NGC 6939}
\shortauthors{Ko{\c c} et al.}
\usepackage{xcolor}
\begin{document}

\title{A photometric and astrometric study of the open clusters NGC 1664 and NGC 6939}

\correspondingauthor{Seliz Ko{\c c}}
\email{seliskoc@gmail.com}

\author[0000-0001-7420-0994]{Seliz Ko{\c c}}
\affiliation{Istanbul University, Institute of Graduate Studies in Science, Programme of Astronomy and Space Sciences, 34116, Beyaz{\i}t, Istanbul, Turkey}

\author[0000-0002-5657-6194]{Talar Yontan}
\affiliation{Istanbul University, Faculty of Science, Department of Astronomy and Space Sciences, 34119, Beyaz\i t, Istanbul, Turkey}

\author[0000-0003-3510-1509]{Sel\c{c}uk Bilir}
\affiliation{Istanbul University, Faculty of Science, Department of Astronomy and Space Sciences, 34119, Beyaz\i t, Istanbul, Turkey}

\author[0000-0003-2575-9892]{Remziye Canbay}
\affiliation{Istanbul University, Institute of Graduate Studies in Science, Programme of Astronomy and Space Sciences, 34116, Beyaz{\i}t, Istanbul, Turkey}

\author[0000-0002-0688-1983]{Tansel Ak}
\affiliation{Istanbul University, Faculty of Science, Department of Astronomy and Space Sciences, 34119, Beyaz\i t, Istanbul, Turkey}

\author[0000-0001-9445-4588]{Timothy Banks}
\affiliation{Nielsen, Data Science, 200 W Jackson Blvd \#17, Chicago, IL 60606, USA}
\affiliation{Physics \& Astronomy, Harper College, 1200 W Algonquin Rd, Palatine, IL 60067, USA}

\author[0000-0002-0912-6016]{Serap Ak}
\affiliation{Istanbul University, Faculty of Science, Department of Astronomy and Space Sciences, 34119, Beyaz\i t, Istanbul, Turkey}

\author[0000-0002-3304-5200]{Ernst Paunzen}
\affiliation{Department of Theoretical Physics and Astrophysics, Masaryk University, Kotl\'a\u rsk\'a 2, 611 37 Brno, Czech Republic}

\begin{abstract}
This study calculated astrophysical parameters, as well as kinematic and galactic orbital parameters, of the open clusters NGC 1664 and NGC 6939.  The work is based on CCD {\it UBV} and {\it Gaia} photometric and astrometric data from ground and space-based observations. Considering {\it Gaia} Early Data Release 3 (EDR3) astrometric data, we determined membership probabilities of stars located in both of the clusters. We used two-color diagrams to determine $E(B-V)$ color excesses for NGC 1664 and NGC 6939 as $0.190 \pm 0.018$ and $0.380 \pm 0.025$ mag, respectively. Photometric metallicities for the two clusters were estimated as [Fe/H] = $-0.10 \pm 0.02$ dex for NGC 1664 and as [Fe/H] = $-0.06 \pm 0.01$ dex for NGC 6939. Using the reddening and metallicity calculated in the study, we obtained distance moduli and ages of the clusters by fitting {\sc parsec} isochrones to the color-magnitude diagrams based on the most likely member stars. Isochrone fitting distances are $1289 \pm 47$ pc and $1716 \pm 87$ pc, which coincide with ages of $675 \pm 50$ Myr and $1.5 \pm 0.2$ Gyr for NGC 1664 and NGC 6939, respectively. We also derived the distances to the clusters using {\it Gaia} trigonometric parallaxes and compared these estimates with the literature. We concluded that the results are in good agreement with those given by the current study. Present day mass function slopes were calculated as $\Gamma=-1.22\pm0.33$ and $\Gamma=-1.18\pm0.21$ for NGC 1664 and NGC 6939, respectively, which are compatible with the \citet{Salpeter55} slope. Analyses showed that both of clusters are dynamically relaxed. The kinematic and dynamic orbital parameters of the clusters were calculated, indicating that the birthplaces of the clusters are outside the solar circle.
 
\end{abstract}

\section{Introduction}
\label{sec:introduction}
Open clusters (OCs) are groupings of stars which formed from the same molecular cloud and so  under similar physical conditions. This results in OCs being gravitationally bound systems and ensures that cluster stars are very similar in distance, age, chemical compositions, positions, and velocities \citep{Lada03}. These characteristics make it possible to determine which of the stars observed in the direction of the cluster are most likely cluster member stars, and then to determine precise astrometric (proper motions and parallaxes), astrophysical (distance, reddening, metallicity, and age) and kinematic (space velocities and orbital motions) parameters of the clusters. Member stars of clusters will have different luminosities within a wide range of stellar masses, according to the initial distribution of masses they formed with.  Studies across OCs can give important information about their dynamical properties as well as insight into the evolution of the Galaxy \citep{Moraux16, Gilmore12, Carraro94}. Such work requires that the cluster member stars should be determined carefully and that the parameters of OCs should be determined precisely using homogeneous data and methods. 

OC parameters have been estimated across many studies and compiled in large catalogs \citep{Sampedro17, Kharchenko13, Dias14, Roeser10, Dias02}. When the distances, color excesses, and ages of OCs in these catalogs are compared there are still inconsistencies across the parameters \citep{Netopil15}.

Using data from the {\it Gaia} mission will shed light on the precise determination of cluster membership as well as being  helpful for astrometric solutions and  parallaxes of OCs. It will also be helpful for the identification of `new' OCs. \citet{Cantat-Gaudin18} used the second data release of {\it Gaia} survey \citep[hereafter Gaia DR2,][]{Gaia18} and determined the membership probabilities of 1,229 OCs with 60 newly detected clusters. Shortly after the third (early) data release of {\it Gaia} \citep[hereafter Gaia EDR3,][]{Gaia21} nearly 865 new galactic OCs were discovered \citep{Castro-Ginard20, Sim19, Liu19}. The {\it Gaia} EDR3 database contains more than 1.55 billion full astrometric solutions and photometric data with high-precision. The trigonometric parallax ($\varpi$) uncertainties are 0.02-0.07 mas for $G\leq17$ mag, reaching 0.5 and 1.3 mas for $G=20$ and $G=21$ mag respectively. The proper motion uncertainties are 0.02~--~0.07 mas yr$^{-1}$ for $G\leq17$, increasing to 0.5 mas yr$^{-1}$ and 1.4 mas yr$^{-1}$ at $G=21$ mag. The $G$-band photometric uncertainties are 0.3~--~1 mmag for the sources within $G\leq17$ mag, climbing to 6 mmag at $G=20$ mag.  The database provides radial velocities for about 7.21 million stars in the $4<G<13$ magnitude range with an accuracy about 1.2 km s$^{-1}$.

In this study we determined the membership probability of cluster stars, mean proper motion, and distance of the OCs NGC~1664 and NGC~6939 using with ground-based {\it UBV} photometric observations together high-precision astrometry and photometry taken from the {\it Gaia} EDR3 catalogue. We present the fundamental parameters, the kinematics and galactic orbital parameters, the luminosity and mass functions, the dynamical state of mass segregation of these clusters. The literature summaries of the clusters are as follows.

% Table 1
\begin{table*}
\setlength{\tabcolsep}{2pt}
\renewcommand{\arraystretch}{0.55}
\small
%\scriptsize
  \centering
  \caption{{\label{tab:table_one} Fundamental parameters for NGC 1664 and NGC 6939 derived in this study and compiled from the literature: columns present cluster names, color excesses ($E(B-V$)), distance moduli ($\mu$), distances ($d$), iron abundances ([Fe/H]), age ($t$), and proper motion components ($\langle\mu_{\alpha}\cos\delta\rangle$, $\langle\mu_{\delta}\rangle$).}}
  \begin{tabular}{cccccccccc}
    \hline
    Cluster &  $E(B-V)$ & $\mu$ & $d$ & [Fe/H] & $t$ &  $\langle\mu_{\alpha}\cos\delta\rangle$ &  $\langle\mu_{\delta}\rangle$ & Ref \\
            & (mag) & (mag) & (pc)  & (dex) & (Myr) & (mas yr$^{-1}$) & (mas yr$^{-1}$) &  \\
    \hline
NGC 1664 & 0.30 & 11.50  & 1300         & ---   & 90         &  ---            & ---              & (1) \\
         & 0.31 & 11.09  & 1240         & ---   & ---        &  ---            & ---              & (2) \\  
         & 0.25 & 11.17  & 1199         & ---   & 525        & -2.54$\pm$0.07  & -6.21$\pm$0.07   & (3) \\  
         & 0.25 & 10.48  & 1200         & ---   & 560$\pm$33 & -4.23$\pm$0.02  & -4.20$\pm$0.02   & (4) \\  
         & ---  & ---    & ---          & ---   & ---        & -1.83$\pm$0.08  & -4.15$\pm$0.17   & (5) \\  
         & ---  & ---    & ---          & -0.11$\pm$0.04 & ---        & ---             &     ---          & (6) \\  
         & 0.25 & 11.17  & 1200         & ---   & 560$\pm$33 & ---             &     ---          & (7) \\           
         & 0.25 & 11.17  & 1200         & ---   & 560        & ---             &     ---          & (8) \\           
         & 0.25 & 11.17  & 1199         & ---   & 290        & -1.36$\pm$0.27  & -5.96$\pm$0.21   & (9) \\           
         & ---  & ---    & 1199         & ---   & 290        & ---             &     ---          & (10) \\  
         & ---  & ---    & 1199$\pm$120 & ---   & 525        & -2.54$\pm$0.31  & -6.21$\pm$0.32   & (11) \\            
         & ---  & ---    & 1362$\pm$7   & ---   & ---        & 1.703$\pm$0.014 & -5.738$\pm$0.010 & (12) \\              
         & ---  & ---    & 1357$\pm$86  & 1.21  & 372$\pm$22 & 1.733$\pm$0.015 & -5.701$\pm$0.012 & (13) \\             
& 0.228$^{+0.006}_{-0.005}$& 10.626$^{+0.020}_{-0.020}$ & 963$^{+9}_{-9}$& ---& 470$^{+20}_{-7}$ & ---  & ---  & (14) \\
         & ---  & ---    & ---          & -0.01$\pm$0.01 & ---        & ---             &     ---          & (15) \\
         & ---  & ---    & ---          & -0.03$\pm$0.01   & 560        & 1.84$\pm$0.07  & -5.76$\pm$0.04   & (16) \\ 
         & 0.28$\pm$0.03 & 11.298$\pm$0.171  & 1218$\pm$49         & -0.009$\pm$0.013   & 645$\pm$100 &1.727$\pm$0.014 & -5.720$\pm$0.010 & (17) \\
         & ---  & ---    & 1302         &  ---     & 525     & ---             &     ---          & (18) \\
         & ---  & ---    & 1351$\pm$57  & ---   & 560$\pm$30 & 1.565$\pm$0.042 & -5.755$\pm$0.068 & (19) \\
         & 0.190$\pm$0.018  & 11.205$\pm$0.075  & 1328$\pm$46         & -0.10$\pm$0.02   & 675$\pm$50        &1.594$\pm$0.071 & -5.780$\pm$0.052 & (20) \\ 
  \hline
NGC 6939 & 0.33$\pm$0.07 & 12.27$\pm$0.41  & 1780$\pm$365 & --- & 1600$\pm$300 & ---  & ---     & (21) \\  
         & 0.34-0.38     & 12.35-12.58     & 1820-1905    & --- & 1000-1300    & ---  & ---     & (22) \\
         & ---           & ---             & ---          & 0.00$\pm$0.10       & ---  & ---     & ---   & (23) \\
& 0.38$^{+0.18}_{-0.10}$ & 12.15$^{+0.56}_{-0.72}$  & 1560$^{+640}_{-590}$ & ---& 1260 & ---     & ---   & (24) \\
         & ---  & ---    & ---             & -0.22$\pm$0.07  & ---                 &   ---       &  ---  & (25) \\ 
         & 0.31 & 12.30  & 1800            & ---   & 1870    &-2.13            & -1.34          & (26) \\ 
         & ---  & ---    & ---             & ---    & ---     &-2.37$\pm$0.09   & -5.29$\pm$0.08 & (5) \\ 
         & 0.31          & 12.24           & 1800   & ---     & 1900            & ---   & ---    & (7) \\
         & 0.33 & 12.30  & 1800            & ---    & 1600    &-2.17$\pm$0.20   & -4.51$\pm$0.19 & (9) \\ 
         & ---  & ---    & 1868$\pm$101    & ---    & ---     &-1.841$\pm$0.006 & -5.413$\pm$0.006 & (12) \\ 
         & ---  & ---    & ---             & 0.098$\pm$0.030    & ---           &---         & --- & (27) \\ 
         & ---  & ---    & 1969$\pm$164    & 1.51    & 1600$\pm$100    &-1.845$\pm$0.007   & -5.408$\pm$0.007 & (13) \\ 
         & 0.30$\pm$0.01 & 12.143$\pm$0.104& 1757$\pm$51 & 0.462$\pm$0.10   & 1600$\pm$200 &-1.838$\pm$0.007 & -5.410$\pm$0.006 & (17) \\
         & ---  & ---    & 1770            &  ---    & 1750    & ---             &     ---          & (18) \\
         & ---  & ---    & 1898$\pm$90     & ---     & 1860 & -1.819$\pm$0.006 & -5.463$\pm$0.006 & (19) \\
         &0.380$\pm$0.025& 12.350$\pm$0.109& 1716$\pm$87    & -0.06$\pm$0.01   & 1500$\pm$200    &-1.817$\pm$0.039   & -5.462$\pm$0.039 & (20) \\ 
           \hline
    \end{tabular}%
    \\
(1) \citet{Lindoff68}, (2) \citet{Fang70}, (3) \citet{Kharchenko05}, (4) \citet{Kharchenko13}, (5) \citet{Dias14}, (6) \citet{Reddy16}, (7) \citet{Joshi16}, (8) \citet{Kharchenko16}, (9) \citet{Sampedro17}, (10) \citet{Zhai17}, (11) \citet{Conrad17}, (12) \citet{Cantat-Gaudin18}, (13) \citet{Liu19}, (14) \citet{Bossini19}, (15) \citet{Carrera19}, (16) \citet{Donor20}, (17) \citet{Dias21}, (18) \citet{Tarricq21}, (19) \citet{Hao21}, (20) This study, (21) \citet{Rosvick02}, (22) \citet {Andreuzzi04}, (23) \citet{Jacobson07}, (24) \citet{Maciejewski07}, (25) \citet{Warren09}, (26) \citet{Kharchenko12}, (27) \citet{Casamiquela19} 

\end{table*}%

%---------------------------------------------------------------

\subsection{NGC 1664}

NGC 1664 ($\alpha=04^{\rm h} 51^{\rm m} 06^{\rm s}$, $\delta= +43^{\rm o} 40^{\rm '} 30^{\rm''}$, $l=161^{\rm o}.68$, $b=-0^{\rm o}.45$) was classified by \citet{Trumpler30} as II2m-a with a tenuous central star concentration. Research literature on the cluster has been based on photoelectric and photographic methods, with no detailed CCD {\it UBV} study prior to the current study. The distance of NGC 1664 was determined by different researchers as being 1100-1300 pc \citep{Becker71, Fang70, Hoag65, Larsson57}. Based on photographic observations, \citet{Larsson57} obtained the color excess of NGC 1664 as $E(B-V)=0.30$ mag while \citet{Lindoff68} determined the age as $t=90$ Myr. \citet{Kharchenko12} applied an algorithm they had developed, for the determination of the astrophysical parameters of clusters taking into account stellar evolution models, to calculate the distance, color excess, and age of NGC 1664 as $d=1200$ pc, $E(B-V)=0.25$ mag, and $t=560$ Myr, respectively, \citep{Kharchenko13}. \citet{Joshi16} determined the structural parameters, age, mass, and color excess of the cluster with a statistical analysis approach. They estimated the color excesses and age of the cluster as $E(B-V)=0.25$ mag and $t=560\pm33$ Myr, respectively. \citet{Reddy16} analyzed high-resolution ($R\sim$ 40,000 to 55,000) spectroscopic observations of red clump stars in 12 OCs, including NGC~1664. They identified two red clump stars as being members of NGC 1664 based on the WEBDA database\footnote{https://webda.physics.muni.cz/} and measured the metallicity of the cluster as ${\rm [Fe/H]}=-0.15\pm0.05$ dex. \citet{Bossini19} evaluated the {\it Gaia} DR2 astrometric and photometric data across 269 OCs, applying a Bayesian statistical analysis to derive the age, distance modulus, and extinction for each of them (see Table 1 for the NGC~1664 results). The authors estimated the distance modulus to be $\mu=10.626\pm0.020$ mag, which corresponds to $d=963\pm9$ pc in distance, and extinction to be $A_{\rm V}=0.708\pm0.018$ mag, which corresponds to a color excess of  $E(B-V)=0.228\pm0.006$ mag. \citet{Carrera19} used APOGEE and GALAH spectroscopic data to estimate the radial velocities of 145 OCs, as well as the iron abundances of 104 OCs. They took into account the two most likely members of NGC 1664 and from these measured the mean radial velocity of the cluster as being $V_{\rm r}= 6.70\pm0.05$ km s$^{-1}$. The iron abundance of the cluster was determined as ${\rm [Fe/H]}=-0.01\pm0.01$ dex from one cluster member star. \citet{Donor20}, using high-resolution spectroscopic data from APOGEE DR16, calculated the abundances across 16 elements for stars estimated as most likely to be cluster members of 128 OCs. Based on {\it Gaia} DR2 astrometric and photometric data membership probabilities of stars, mean proper motion components and cluster ages were determined. Iron abundance was measured as ${\rm [Fe/H]}=-0.03\pm0.01$ dex for NGC~1664, based on high-resolution spectroscopic data of one cluster member star. \citet{Donor20} estimated the age of the cluster as $t=560$ Myr and the mean proper motions components as ($\mu_{\alpha}\cos\delta$, $\mu_{\delta}$)=($1.84\pm0.07, -5.76\pm0.04$) mas yr$^{-1}$.

%---------------------------------------------------------------

\subsection{NGC 6939}

NGC 6939 ($\alpha=20^{\rm h} 31^{\rm m} 30^{\rm s}$, $\delta= +60^{\rm o} 39^{\rm '} 42^{\rm''}$, $l=95^{\rm o}.90$, $b=+12^{\rm o}.30$) is a relatively star-rich cluster located in the constellation Cepheus, close to the galactic plane. The first study of the cluster was made by \citet{Kustner23}, providing photographic data of 370 stars detected in the cluster vicinity. \citet{Cuffey44} determined the distance of the cluster as $d=1300$ pc, while \citet{Cannon69} estimated its age as being approximately $t=1$ Gyr. \citet{Chincarini63} identified likely cluster member stars up to $V\sim 15.5$ mag through analysing photoelectric and photographic data. The study estimated that the color excess, distance modulus, and age of NGC 6939 as $E(B-V)=0.5$ mag, $\mu=12.00$ mag, and $t=50$ Myr respectively. Later researchers obtained photoelectric and radial velocities measurements, together with studies of the proper motion components of likely cluster member stars \citep{Milone94, Mermilliod94, Glushkova91, Geisler88}. \citet{Robb98} utilized their CCD observations to detect six variable stars in the cluster region, and classified three of them as K giant variables and the others as eclipsing binary systems from the variation of their long-term light curves. The first CCD photometric study of the NGC 6939 was made by \citet{Rosvick02},  analysing {\it BVI} data for stars situated near the cluster center. \citet{Rosvick02} determined the age of cluster as $t=1.6\pm0.3$ Gyr and remarked that color excess of NGC 6939 changes across the body of the cluster within the range $0.29 < E(B-V) < 0.41$ mag. 

Following this, \citet{Andreuzzi04} performed a CCD {\it UBVI} photometric study of NGC 6939 and determined its distance as $d=1800$ pc. They estimated that age of the cluster was in the range $t$=1.6 to 1.3 Gyr and that the color excess across the cluster changes within the range $0.34 < E(B-V) < 0.38$ mag. \citet{Maciejewski07} undertook a CCD {\it BV} photometric analysis of 42 OCs including NGC 6939 and determined the age, color excess, distance modulus, and distance of this cluster as being $t=1.25$ Gyr, $E(B-V)=0.38^{+0.18}_{-0.10}$ mag, $\mu=12.15^{+0.56}_{-0.72}$ mag, and $d=1560^{+640}_{-590}$ pc respectively.

The structural analysis of \citet{Maciejewski07} determined the core radius of cluster to be $r_{c}=2.2$ arcmin. Based on mass function results, they indicated that there were `evaporated' (or escaped) stars from the cluster. \citet{Kharchenko12} took into account stellar evolution models in their estimation of the distance, reddening, and age of the cluster as being $d=1800$ pc, $E(B-V)=0.31$ mag, and $t=1.9$ Gyr respectively (see Table 1 for the NGC~1664 results). 

Considering the literature, it can be seen that NGC 6939 is an intermediate-age open cluster. This implies the existence of evolved giant stars in the cluster, whose presence has enabled researchers to determine the metallicity of the cluster via photometric and spectroscopic data \citep{Casamiquela17, Dias14, Jacobson13, Jacobson07, Friel02, Thogersen93, Geisler91, Canterna86}.  Across these studies, the iron abundance is within $-0.14\leq {\rm  [Fe/H]}\leq 0.13$ dex. Difference of metallicity estimates is likely due to different observation methods and analysis techniques used by the studies. 

% FIGURE 1
\begin{figure*}
\centering
\includegraphics[scale=0.52, angle=0]{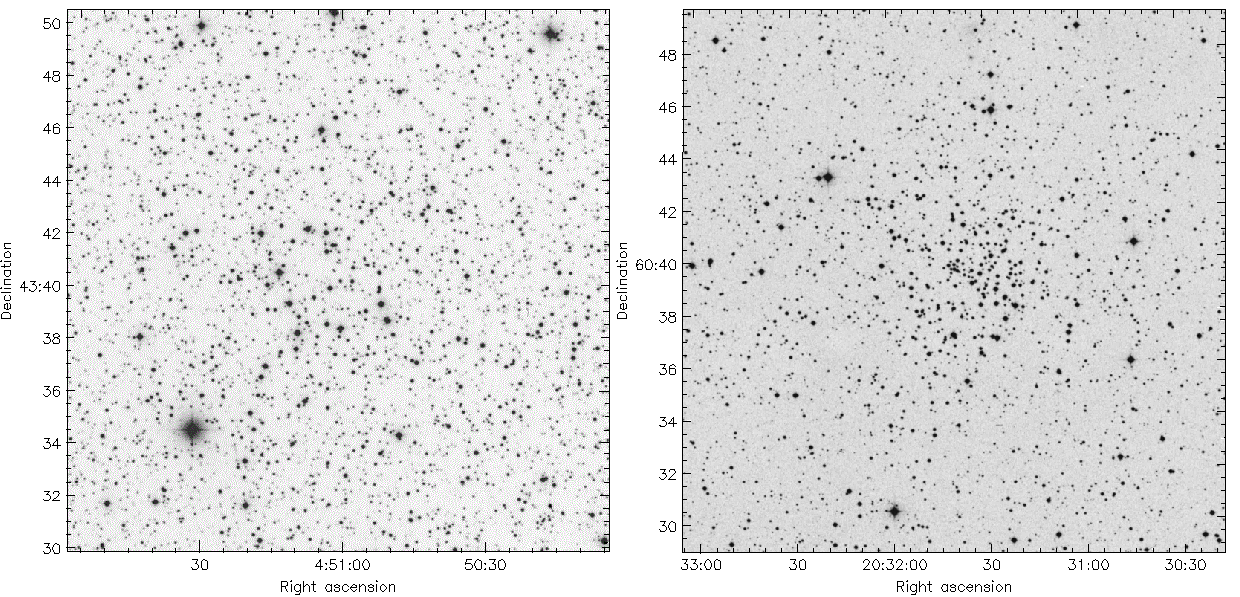}
\caption{Identification chart of stars in NGC 1664 (left panel) and for NGC 6939 (right panel), taken from Leicester database and archive service (LEDAS).} 
\end {figure*}

%---------------------------------------------------------------

\section{Observations}

The ground-based CCD {\it UBV} observations of NGC 1664 and NGC 6939 were made using the 100-cm Ritchey-Chr\'etien telescope (T100) of the T\"UB\.ITAK  National Observatory (TUG)\footnote{www.tug.tubitak.gov.tr} located in Turkey. Identification maps of the open clusters NGC 1664 and NGC 6939 are shown in Fig. 1. The {\it UBV} images were taken using a Spectral Instruments CCD camera operating at $-100^{\rm o} \rm{C}$. The CCD camera is equipped with a back illuminated 4k$\times$4k pixel CCD, with a scale of $0''.31$ pixel$^{-1}$. This results in an unvignetted field of view of $21'.5 \times 21'.5$ on the telescope. The gain and the readout noise of the CCD camera are 0.55 e$^{-}$/ADU and 4.19~e$^{-}$ (100 KHz, Ch:A), respectively. In order to be sensitive to the widest possible flux range (from UV and IR), different exposure times were used during the observations. A log of observations is given in Table~\ref{tab:exposures}. $UBV$ photometric calibrations were based on \citet{Landolt09} standard stars. A total of 108 stars in 15 selected areas were observed. The air masses were between 1.232 and 1.918 for the standard stars monitored during the observation nights (see Table~\ref{tab:standard_stars}). Standard CCD reduction techniques were applied using IRAF\footnote{IRAF is distributed by the National Optical Astronomy Observatories} packages for the pre-processing of all the images which were taken. Astrometric corrections were applied using PyRAF\footnote{PyRAF is a product of the Space Telescope Science Institute, which is operated by AURA for NASA} and astrometry.net\footnote{http://astrometry.net} routines together with our own scripts for the cluster images. Using IRAF's aperture photometry packages, we measured the instrumental magnitudes of \citet{Landolt09} stars. We next applied multiple linear fits to these magnitudes to derive photometric extinction and transformation coefficients for each observing night as listed in Table~\ref{tab:coefficients}. Source Extractor (SExTractor) and PSF Extractor (PSFEx) routines \citep{Bertin96} were applied to measure the instrumental magnitudes of the objects located in the cluster fields. We applied aperture corrections to these magnitudes and transformed them to standard magnitudes in the Johnson photometric system using transformation equations as described by \citet{Janes11}: 

% Table 2
\begin{table}
\setlength{\tabcolsep}{4.5pt}
  \centering
  \caption{Observation log:  columns denote name of clusters, observation date, filters, exposure times (in seconds), and the number of exposures ($N$). Dates are day-month-year.}
    \begin{tabular}{cclll}
    \hline
            &  & \multicolumn{3}{c}{Filter/Exp, Time (s) $\times N$} \\
    \hline
    Cluster & Obs. Date & $U$ & $B$ & $V$ \\
    \hline
    NGC 1664 & 05-11-2018 &   60$\times$2 & 6$\times$3   &   4$\times$5 \\
             &            & 1800$\times$2 & 600$\times$2 & 300$\times$3 \\
    NGC 6939 & 30-07-2019 & 300$\times$3 & 40$\times$5  &  15$\times$5 \\
             &            & 1200$\times$1 & 900$\times$2 & 600$\times$2 \\
    \hline
    \end{tabular}%
  \label{tab:exposures}%
\end{table}%

% Table 3
\begin{table}
\setlength{\tabcolsep}{3pt}
\renewcommand{\arraystretch}{0.6}
  \centering
  \caption{Information on the observations of standard stars from selected \citet{Landolt09} fields. The columns are the observation date, star field name as from Landolt, the number of standard stars ($N_{\rm st}$) observed in a given field, the number of pointings to each field ($N_{\rm obs}$, i.e., observations), and the airmass range the fields (on a given night) were observed over ($X$). Dates are day-month-year.}

    \begin{tabular}{llccc}
    \hline
Date	   & Star Field	& $N_{\rm st}$	& $N_{\rm obs}$	& $X$\\
\hline
           & SA92SF2    & 15            & 1	            &          \\
	       & SA93       &  4	        & 1	            &          \\
	       & SA94	    &  2 	        & 1	            &          \\
           & SA95SF2    &  9	        & 1	            &          \\
05-11-2018 & SA96	    &  2	        & 2	            & 1.232 -- 1.918 \\
	       & SA97SF1    &  2	        & 2	            &          \\
	       & SA99	    &  3	        & 1	            &          \\
	       & SA100SF2   & 10	        & 1	            &          \\
	       & SA112	    &  6	        & 1	            &          \\
	       & SA114	    &  5	        & 1	            &          \\	       
  \hline	      
      	   & SA92SF2    & 15 	        & 1	            &          \\
      	   & SA93       &  4	        & 1	            &          \\
      	   & SA94	    &  2	        & 1	            &          \\
      	   & SA106      &  2	        & 1	            &          \\
30-07-2019 & SA107	    &  7	        & 1	            & 1.229 -- 1.892\\
           & SA108	    &  2	        & 1	            &          \\
           & SA109	    &  2	        & 1	            &          \\
	       & SA111	    &  5	        & 1	            &          \\
	       & SA112	    &  6	        & 2	            &          \\
	       & SA114	    &  5	        & 1	            &          \\
    \hline
    \end{tabular}%
  \label{tab:standard_stars}%
\end{table}%

% Table 4 --- 
\begin{table*}
\renewcommand{\arraystretch}{0.6}
  \centering
  \caption{Transformation and extinction coefficients derived for the two observation nights: $k$ and $k'$ are the primary and secondary extinction coefficients, respectively. $\alpha$ and $C$ are the transformation coefficients. Dates are day-month-year.} 
    \begin{tabular}{lccccc}
    \hline
    Filter/color index & Obs. Date       & $k$     & $k'$    & $\alpha$     & $C$ \\
    \hline
$U$     & 05.11.2018 & 0.502$\pm$0.078 & $-0.117\pm$0.095 & ---               & --- \\
$B$     &            & 0.232$\pm$0.046 & $-0.037\pm$0.050 & 0.968$\pm$0.076 & 1.488$\pm$0.070 \\
$V$     &            & 0.128$\pm$0.020 & ---              & ---               & --- \\
$U-B$   &            & ---             & ---              & 0.977$\pm$0.143 & 3.724$\pm$0.121 \\
$B-V$   &            & ---             & ---              & 0.078$\pm$0.009 & 1.550$\pm$0.031 \\
$U$     & 30.07.2019 & 0.498$\pm$0.092 & $-0.068\pm$0.108 & ---               & --- \\
$B$     &            & 0.293$\pm$0.079 & $-0.026\pm$0.082 & 0.866$\pm$0.127 & 1.799$\pm$0.116 \\
$V$     &            & 0.208$\pm$0.026 & ---              & ---               & --- \\
$U-B$   &            & ---             & ---              & 0.953$\pm$0.161 & 4.162$\pm$0.130 \\
$B-V$   &            & ---             & ---              & 0.073$\pm$0.012 & 1.782$\pm$0.037 \\
\hline
    \end{tabular}%
  \label{tab:coefficients}%
\end{table*}%
 
\begin{equation}
V = v - \alpha_{bv}(B-V)-k_vX _v- C_{bv} \\
\end{equation}

\begin{equation}
B-V = \frac{(b-v)-(k_b-k_v)X_{bv}-(C_b-C_{bv})}{\alpha_b+k'_bX_b-\alpha_{bv}} \\
\end{equation}

\begin{eqnarray}
U-B = \frac{(u-b)-(1-\alpha_b-k'_bX_b)(B-V)}{\alpha_{ub}+k'_uX_u}-\frac{(k_u-k_b)X_{ub}-(C_{ub}-C_b)}{\alpha_{ub}+k'_uX_u} 
\end{eqnarray}
Here $U$, $B$, and $V$ indicate the magnitudes in the standard photometric 
system. $u$, $b$, and $v$ are the instrumental magnitudes. $X$ is the airmass. 
$k$ and $k'$ are the primary and secondary extinction coefficients while 
$\alpha$ and $C$ are transformation coefficients to the standard system. 
The photometric extinction and transformation coefficients for a
given night were calculated by applying multiple linear fits to the 
instrumental magnitudes of the standard stars. The obtained values are 
given in Table~\ref{tab:coefficients}.

% TABLE 5
\begin{table*}
\setlength{\tabcolsep}{1.9pt}
\renewcommand{\arraystretch}{0.8}
%\scriptsize
\tiny
  \centering
  \caption{\label{tab:input_parameters}
The catalogues for NGC 1664 and NGC 6939. The complete table can be found in the online version of this article}.
    \begin{tabular}{cccccccccccc}
\hline
\multicolumn{11}{c}{NGC 1664}\\
\hline
ID	 & RA           &	DEC	        &      $V$	    &	$U-B$      & $B-V$	      &	$G$	          & 	$G_{\rm BP}-G_{\rm RP}$	 & 	$\mu_{\alpha}\cos\delta$ & 	$\mu_{\delta}$ & 	$\varpi$	& $P$ \\

	 & (hh:mm:ss.ss)           &	(dd:mm:ss.ss)	&      (mag)	    &	(mag)      & (mag)	      &	(mag)	          & 	(mag)	 & 	(mas yr$^{-1}$) & 	(mas yr$^{-1}$) & 	(mas)	&  \\
\hline
0001 & 04:50:03.99	& +43:40:40.33	& 18.924(0.015) & ---		   & 1.244(0.025) & 18.359(0.003) & 1.626(0.030) & 0.417(0.311) & -1.427(0.196) & 0.375(0.195) & 0.00\\
0002 & 04:50:04.03	& +43:31:37.23	& 19.752(0.038) & ---	       & 1.436(0.064) & 18.712(0.003) & 1.905(0.039) & 1.631(0.319) & -10.150(0.206) & 1.132(0.226) & 0.00\\
0003 & 04:50:04.11	& +43:35:00.27	& 20.025(0.044) & ---          & 1.996(0.106) & 18.495(0.003) & 2.676(0.060) &-2.182(0.278) & -2.044(0.177) & 2.356(0.186) & 0.00\\
0004 & 04:50:04.12	& +43:39:01.56	& 19.264(0.020) & 0.424(0.073) & 1.007(0.030) & 18.833(0.003) & 1.389(0.055) & 2.572(0.373) &	-3.090(0.227) & 0.272(0.217) & 0.00\\
0005 & 04:50:04.14	& +43:45:50.34	& 17.434(0.005) & 0.227(0.013) & 0.735(0.007) & 17.105(0.003) & 1.080(0.011) & 0.594(0.122) & -0.425(0.088) & 0.389(0.083) & 0.00\\
... & ... & ...	& ... & ... & ... & ... & ... & ... & ... & ... & ...\\
... & ... & ...	& ... & ... & ... & ... & ... & ... & ... & ... & ...\\
... & ... & ...	& ... & ... & ... & ... & ... & ... & ... & ... & ...\\
3731 & 04:51:58.31 & +43:46:24.34 & 18.438(0.010) &	0.409(0.034) & 0.975(0.015) & 18.050(0.003) &	1.297(0.019) &  1.281(0.188) &	-1.799(0.156) &	0.127(0.152) & 0.00\\	
3732 & 04:51:58.31& +43:45:56.87 & 19.354(0.025) & --- & 3.249(0.141) &	19.217(0.005) &	2.921(0.105) & -0.487(0.481)	& - 2.882(0.394) & 	1.554(0.422) &	0.00\\	
3733 & 04:51:58.36 & +43:49:50.38 & 17.440(0.005) &	0.509(0.017) &	0.944(0.008) &	17.062(0.003) &	1.234(0.009) &	0.036(0.089) &	-2.873(0.076) & 0.529(0.077) &	0.00\\	
3734 & 04:51:58.38 & +43:40:40.03 & 19.033(0.016) &	0.474(0.056) & 0.996(0.023) & 18.505(0.003) &	1.430(0.029) &	0.597(0.238) &	-2.226(0.191) & 0.122(0.218) &	0.00\\	
3735 & 04:51:58.39 & +43:43:43.06 &	20.054(0.038) & --- & 1.131(0.058) & 19.352(0.004) & 1.474(0.075)	& 0.195(0.539) & -2.410(0.396) & -0.368(0.422) & 0.00\\	
\hline
\multicolumn{11}{c}{NGC 6939}\\
\hline
ID	 & RA           &	DEC	        &      $V$	    &	$U-B$      & $B-V$	      &	$G$	          & 	$G_{\rm BP}-G_{\rm RP}$	 & 	$\mu_{\alpha}\cos\delta$ & 	$\mu_{\delta}$ & 	$\varpi$	& $P$ \\

	 & (hh:mm:ss.ss)           &	(dd:mm:ss.ss)	&      (mag)	    &	(mag)      & (mag)	      &	(mag)	          & 	(mag)	 & 	(mas yr$^{-1}$) & 	(mas yr$^{-1}$) & 	(mas)	&  \\
\hline
0001 & 20:30:04.44 & +60:34:33.13 & 18.664(0.018) & 	--- &  +1.603(0.033) & 18.313(0.010) & +1.368(0.052)  & -1.073(0.130) &  -1.300(0.128) & 0.207(0.103) & 0.00\\
0002 & 20:30:05.22 & +60:45:23.12 &	18.329(0.012) &  0.619(0.109) & 1.091(0.018) & 17.999(0.003) & 	1.324(0.026) & -7.536(0.121) & 0.386(0.114) & 0.280(0.091) & 0.00\\
0003 & 20:30:05.67 & +60:49:43.02 & 18.205(0.011) & 0.857(0.140) & 1.248(0.017) &  17.825(0.003) & 	1.426(0.019) &  -2.033(0.110) & -3.281(0.102) & 0.200(0.081) & 0.00\\
0004 & 20:30:05.71 & +60:36:58.11 & 17.745(0.007) & 0.536(0.051) & 0.879(0.010) & 17.492(0.003) & 	1.145(0.012) & -4.928(0.087) & -5.176(0.084) & 0.357(0.068) & 0.00\\
0005 & 20:30:05.76 & +60:41:56.04 & 19.027(0.022) & --- & 1.067(0.032) & 18.652(0.003) & 1.417(0.034) & -4.526(0.204) & -4.175(0.170) & 0.221(0.151) & 0.00\\
... & ... & ...	& ... & ... & ... & ... & ... & ... & ... & ... & ...\\
... & ... & ...	& ... & ... & ... & ... & ... & ... & ... & ... & ...\\
... & ... & ...	& ... & ... & ... & ... & ... & ... & ... & ... & ...\\

2115 & 20:32:52.50 & +60:41:20.72 & 18.967(0.020) & --- & 1.377(0.034) & 18.553(0.003) & 1.494(0.037) & -0.399(0.167) & -1.332(0.160) & 0.431(0.127) & 0.00\\
2116 & 20:32:52.82 & +60:45:20.23 & 16.327(0.003) & +0.371(0.018) & 0.977(0.004) & 16.022(0.003) & 	1.269(0.006) & 	0.575(0.044) & 0.238(0.039) & 0.359(0.032) & 0.00\\
2117 & 20:32:52.90 & +60:42:07.39 & 18.796(0.019) & --- & 1.825(0.038) & 17.906(0.003) & 2.224(0.021)	& 02.567(0.117) & -9.291(0.111)	& 1.446(0.086) & 0.00\\
2118 & 20:32:52.90 & +60:43:20.38 & 18.689(0.016) & --- & 1.355(0.026) & 18.190(0.003) & 1.647(0.031) & 01.537(0.151)	& 1.456(0.137) &  0.475(0.112) & 0.00\\
2119 & 20:32:53.78 & +60:49:42.70 & 18.898(0.023) & --- & 1.372(0.038) & 18.410(0.003) & 1.629(0.026) & 0.105(0.136) & -5.760(0.142) & 0.575(0.116) & 0.00\\
\hline
    \end{tabular}
      \label{tab:all_cat}%
\end{table*} 

% Table 6
\begin{table*}
\setlength{\tabcolsep}{3pt}
\renewcommand{\arraystretch}{0.8}
  \centering
  \caption{The mean internal photometric errors and number of measured stars in the corresponding $V$ apparent-magnitude interval for each cluster.}
    \begin{tabular}{ccccccc|ccccccc}
      \hline
    \multicolumn{7}{c}{NGC 1664} & \multicolumn{6}{c}{NGC 6939} \\
    \hline
  $V$ & $N$ & $\sigma_{\rm V}$ & $\sigma_{\rm U-B}$ & $\sigma_{\rm B-V}$ & $\sigma_{\rm G}$ &  $\sigma_{G_{\rm BP}-G_{\rm RP}}$ & $V$ & $N$ & $\sigma_{\rm V}$ & $\sigma_{\rm U-B}$ & $\sigma_{\rm B-V}$ & $\sigma_{\rm G}$ & $\sigma_{G_{\rm BP}-G_{\rm RP}}$\\
  \hline
  (08, 12] &  28   & 0.002 & 0.004 & 0.002 & 0.003 & 0.007 & (08, 12] &   6 & 0.001 & 0.002 & 0.001 & 0.003 & 0.005\\
  (12, 14] & 111   & 0.003 & 0.005 & 0.005 & 0.003 & 0.005 & (12, 14] &  73 & 0.002 & 0.005 & 0.002 & 0.003 & 0.005\\
  (14, 15] & 116   & 0.001 & 0.003 & 0.002 & 0.003 & 0.006 & (14, 15] & 123 & 0.001 & 0.006 & 0.002 & 0.003 & 0.005\\
  (15, 16] & 183   & 0.002 & 0.005 & 0.002 & 0.003 & 0.006 & (15, 16] & 172 & 0.002 & 0.013 & 0.002 & 0.003 & 0.005\\
  (16, 17] & 335   & 0.003 & 0.010 & 0.004 & 0.003 & 0.007 & (16, 17] & 263 & 0.003 & 0.026 & 0.005 & 0.003 & 0.007\\
  (17, 18] & 574   & 0.005 & 0.019 & 0.008 & 0.003 & 0.012 & (17, 18] & 339 & 0.006 & 0.064 & 0.010 & 0.003 & 0.011\\
  (18, 19] & 847   & 0.010 & 0.044 & 0.016 & 0.003 & 0.024 & (18, 19] & 506 & 0.013 & 0.102 & 0.023 & 0.003 & 0.025\\
  (19, 20] & 1017  & 0.023 & 0.074 & 0.037 & 0.003 & 0.032 & (19, 20] & 589 & 0.033 & 0.112 & 0.060 & 0.003 & 0.046\\
  (20, 22] & 522   & 0.045 & 0.121 & 0.079 & 0.005 & 0.086 & (20, 22] &  47 & 0.086 & 0.116 & 0.155 & 0.008 & 0.084\\
      \hline
    \end{tabular}%
  \label{tab:phiotometric_errors}%
\end{table*}%

%---------------------------------------------------------------

% FIGURE 2
\begin{figure*}
\centering
\includegraphics[scale=0.55, angle=0]{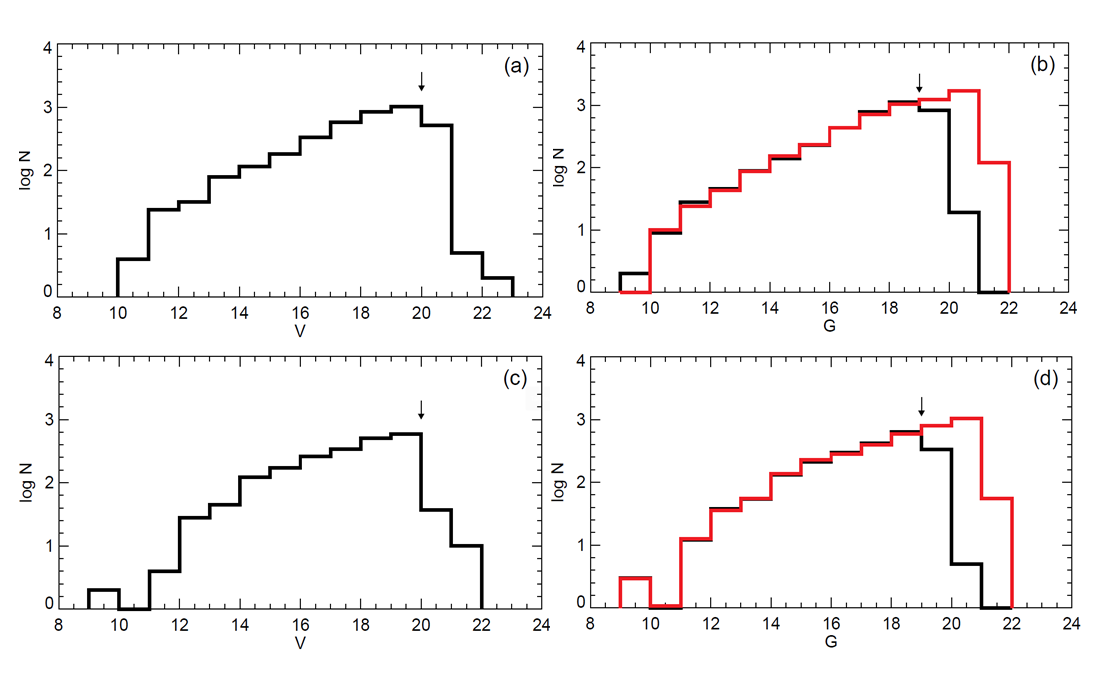}\\
\caption{Interval $V$ and $G$-band magnitude histograms of NGC 1664 (a, b) and NGC 6939 (c, d): The arrows show the faint limiting apparent magnitudes in $V$ and $G$-bands. Black lines indicate the star counts based on the stars detected in the study, while red lines are star counts based on the stars taken from {\it Gaia} EDR3 for the same cluster regions.} 
\label{fig:histograms}
\end {figure*} 

\section{Data Analysis}

\subsection{Photometric data}

Catalogues of ground-based CCD {\it UBV} photometric information listing all the detected stars in the cluster regions are available electronically for NGC 1664 and NGC 6939\footnote{The complete tables can be obtained from VizieR electronically}. We identified 3,735 stars for NGC 1664 and 2,119 stars for NGC 6939, before constructing their photometric and astrometric catalogues. Both of these catalogues contain positions ($\alpha, \delta$), apparent $V$ magnitudes, color indices ($U-B$, $B-V$), proper motion components ($\mu_{\alpha}\cos\delta, \mu_{\delta}$) along with trigonometric parallaxes ($\varpi$) from the {\it Gaia} EDR3, and membership probabilities ($P$) as calculated in this study (Table~\ref{tab:all_cat}). Magnitude and color inaccuracies of Johnson ($V$, $U-B$, $B-V$) and {\it Gaia} EDR3 ($G$, $G_{\rm BP}-G_{\rm RP}$) photometry were adopted as internal errors. We calculated mean photometric errors as a function of $V$ magnitude intervals in Table~\ref{tab:phiotometric_errors}. It can be seen from this table that the mean internal {\it UBV} errors for stars brighter than $V=22$ mag for NGC 1664 and NGC 6939 are smaller than 0.045 mag and 0.086 mag, respectively. Moreover the mean internal errors in the {\it Gaia} EDR3 photometric data for the stars brighter than $V=22$ mag are smaller than 0.005 and 0.008 mag for NGC 1664 and NGC 6939, respectively.

$V$ and $G$ magnitude histograms were used to estimate the photometric completeness limits for each of the clusters (see Fig.~\ref{fig:histograms}). We compared these stellar counts with {\it Gaia} EDR3 data from areas of the same size as the cluster CCD images, using the central equatorial coordinates given by \citet{Cantat-Gaudin20} and utilizing the stars within $8<G<23$ mag range. These distributions are shown in Fig.~\ref{fig:histograms}: Black solid lines represent the observational values according to the $V$ and $G$ interval magnitudes, while the red solid lines (see in Fig.~\ref{fig:histograms}b and \ref{fig:histograms}d) show the stellar distributions retrieved from {\it Gaia} EDR3. Figures~\ref{fig:histograms}b and \ref{fig:histograms}d demonstrate the clear alignment of the stellar counts from these two data sets up to the magnitudes adopted as being the completeness limits (indicated by the vertical arrows in the diagrams) for the clusters. These values are $V=20$ mag for each cluster which corresponds to $G=19$ mag both for NGC 1664 and NGC 6939. Telescope detection limits and telescope-detector combinations vary for ground and space based observations. These differences will lead to the number of detected stars at fainter magnitudes varying between different observational studies. This situation could clarify why the numbers of detected stars at $G>20$ mag in the {\it Gaia} observations are greater than from this study's ground-based photometry. Detected stars are significant for precise analyses. As it can be seen from the Fig.~\ref{fig:histograms}, beyond the completeness limits there are incomplete data. Therefore in analyses we considered the stars brighter than $V=20$ mag for two clusters.

\begin{figure}
\centering
\includegraphics[scale=0.40, angle=0]{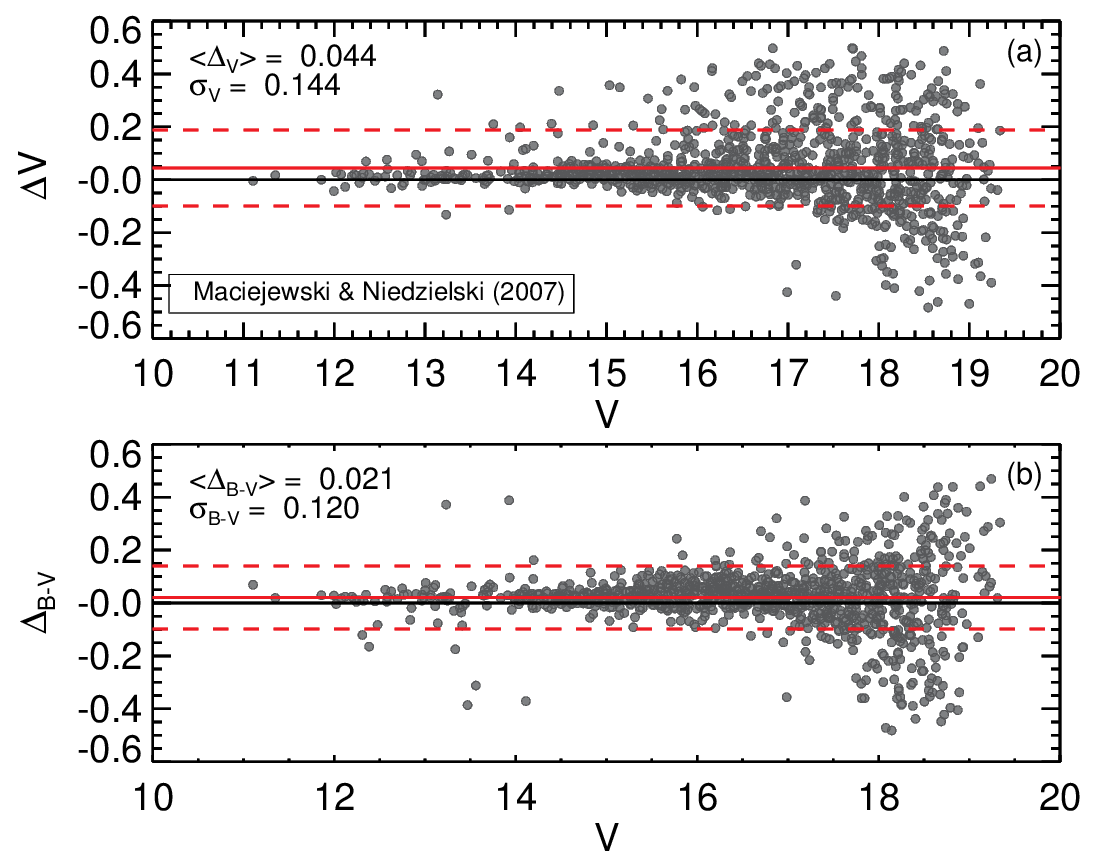}\\
\caption{Comparisons of observational magnitude and colors with those calculated from \citet{Maciejewski07} (a-b) and for NGC 6939. The mean differences and their $\pm 1\sigma$ dispersions are represented with solid and dashed red lines, respectively}
\label{fig:comparison}
\end {figure}

Due to the lack of literature CCD {\it UBV} data for NGC 1664, we compared our photometric results only for NGC 6939 where CCD {\it BV} data were given by \citet{Maciejewski07}. There are 3,129 stars brighter than $V\sim 21$ mag in the catalog of \citet{Maciejewski07}, which we attempted to cross-match with the current study using the stellar equatorial coordinates. 1,150 stars within the $11<V<20$ mag range were successfully matched between the two studies, allowing comparison of the $V$ magnitudes and ($B-V$) colors measurements (see Fig.~\ref{fig:comparison}). The horizontal and vertical axes of this figure denote our $V$ band measurements and differences between $V$ magnitudes and $(B-V)$ colors between the two studies, respectively. Based on these comparisons, the mean differences of magnitude and colors were calculated as $\langle \Delta V\rangle=0.044\pm 0.144$ and $\langle \Delta (B-V)\rangle=0.021\pm 0.120$ mag. We therefore find that the current study's photometry is compatible with that of \citet{Maciejewski07}.    

% FIGURE 3

%---------------------------------------------------------------
\newpage

\subsection{Spatial Structure of the Clusters}

\begin{figure}
\centering
\includegraphics[scale=0.5, angle=0]{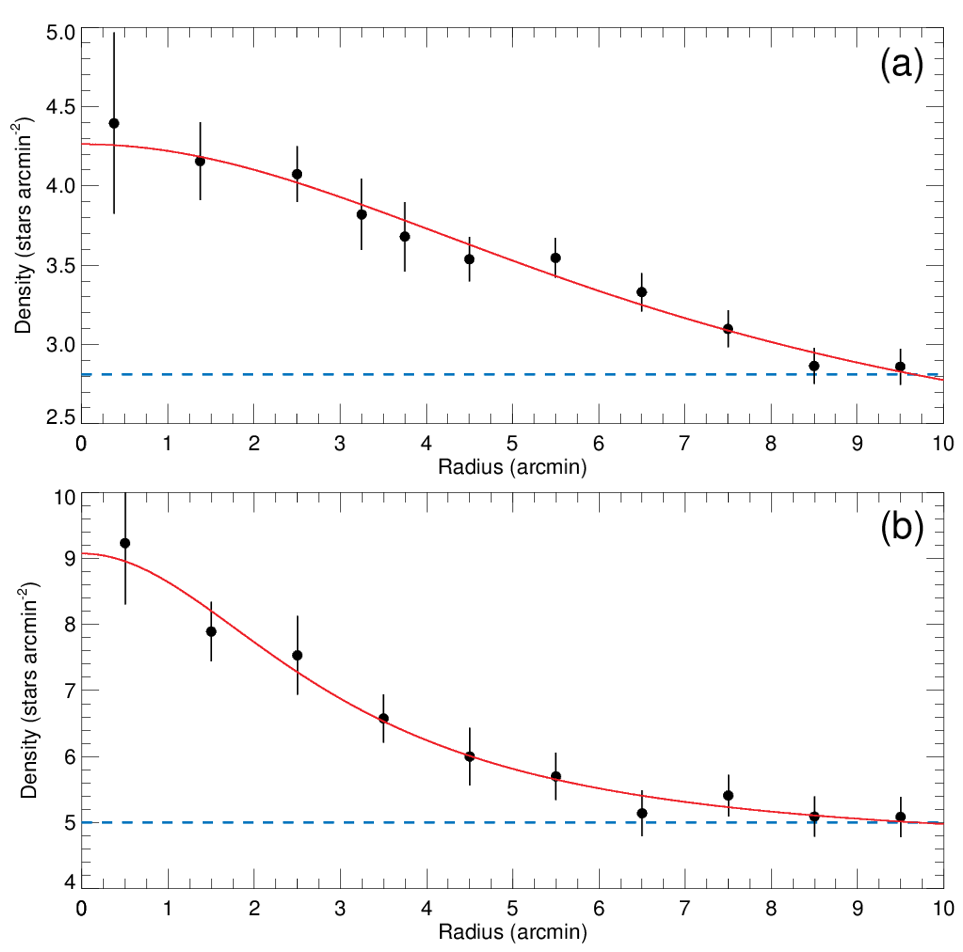}\\
\caption{The radial density profiles for the NGC 1664 (a) and NGC 6939 (b) clusters. Errors were derived using equation of $1/\sqrt N$, where $N$ represents the number of stars used in the density estimation. The fitted red curve represents the \citet{King62} profile. The horizontal blue dashed line shows background stellar density.} 
\label{fig:king}
\end {figure}

We applied the \citet{King62} radial density profile (RDP) model to estimate the core, limiting, and effective radii of NGC 1664 and NGC 6939.  To do this we constructed a series of concentric rings oriented on the central coordinates given by \citet{Cantat-Gaudin20}. Stellar densities ($\rho$) were estimated for each ring by dividing the number of stars found in each ring by the ring area. Figure~\ref{fig:king} plots these measured stellar densities versus radius from the cluster region together with optimal \citet{King62} RDP following $\chi^2$ minimisation. The \citet{King62} model is given by the equation $\rho(r)=f_{\rm bg}+[f_{\rm 0}/(1+(r/r_{\rm c})^2)] $ where $r$ is the radius from the cluster centre, $f_{\rm 0}$ is the central density, $r_{\rm c}$ is the core radius, and $f_{\rm bg}$ is the background density. From \citet{King62} model fit, the central stellar density and core radius of the clusters, together with the background stellar density were inferred as $f_{\rm 0}=1.451\pm 0.281$ stars arcmin$^{-2}$, $r_{\rm c}=7.213\pm 1.696$ arcmin and $f_{\rm bg}=2.812\pm 0.283$ stars arcmin$^{-2}$ for NGC 1664 and $f_{\rm 0}=4.070\pm 0.249$ stars arcmin$^{-2}$, $r_{\rm c}=3.057\pm 0.637$ arcmin and $f_{\rm bg}=5.002\pm 0.328$ stars arcmin$^{-2}$ for NGC 6939, respectively. As seen in Fig.~\ref{fig:king}, the stars begin to merge with the background density (blue dashed lines) at the limiting radius. We therefore concluded that NGC 1664 and NGC 6939 have limiting radii of $r_{\rm lim}=8.5$ ($3.28$ pc) and $r_{\rm lim}=6.5$ arcmin ($3.25$ pc), respectively. In the subsequent analyses for both clusters we considered only the stars inside these limiting radii.

% FIGURE 4

% FIGURE 5
\begin{figure*}
\centering
\includegraphics[scale=0.8, angle=0]{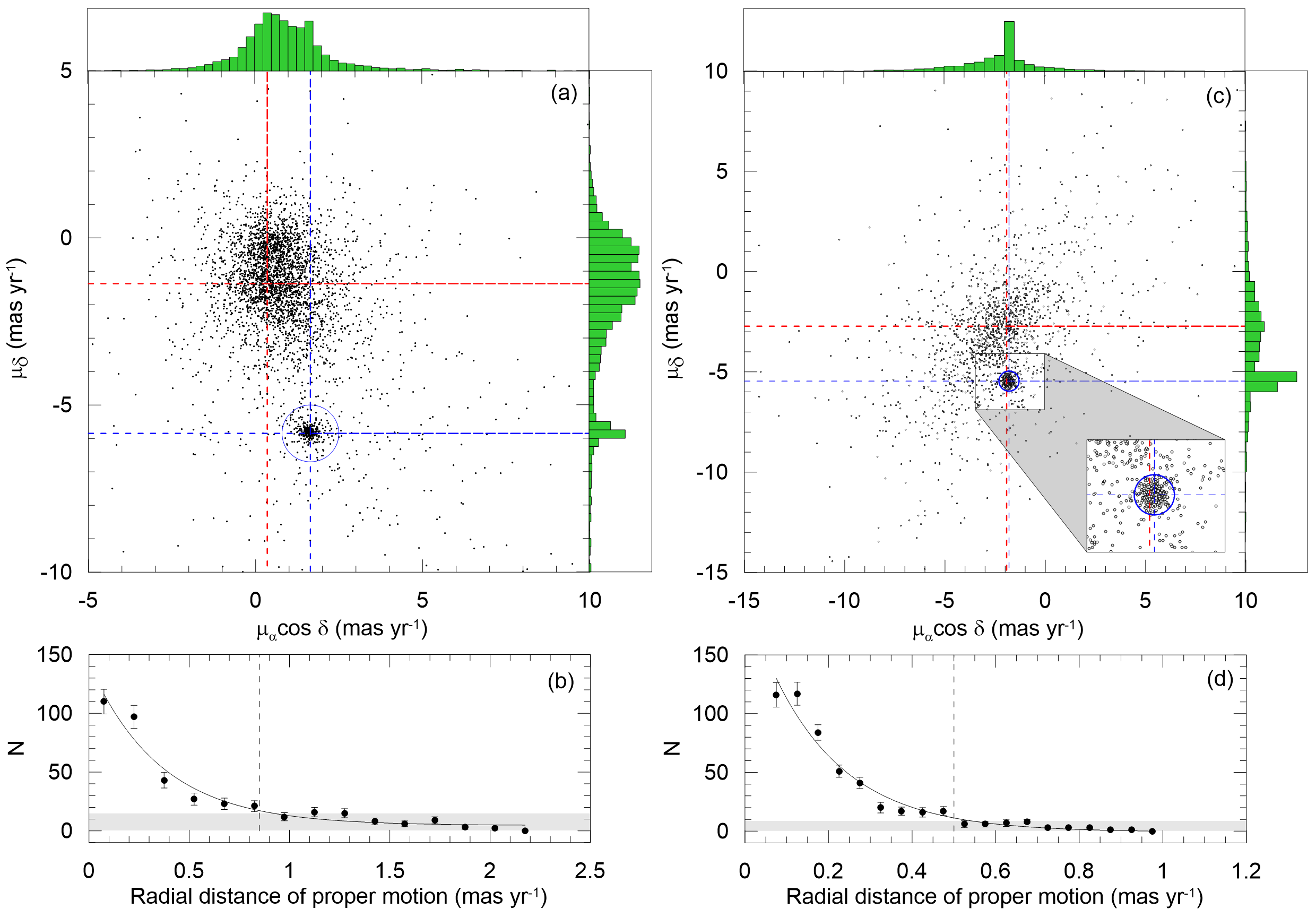}\\
\caption{VPDs and the radial distribution of proper motion diagrams for NGC 1664 (panels a and b) and NGC 6939 (panels c and d).
The zoomed box in panel (c) shows the region of condensation for NGC 6939 in the VPD. The continuous curves in panels (b) and (d) represent King profiles, which are fitted to the data points. Grey horizontal lines represent the field star contamination and dotted lines indicate the radius that adopted to find out the possible cluster members.}
\label{fig:VPD_all}
\end {figure*} 

% FIGURE 6
\begin{figure*}
\centering
\includegraphics[scale=0.54, angle=0]{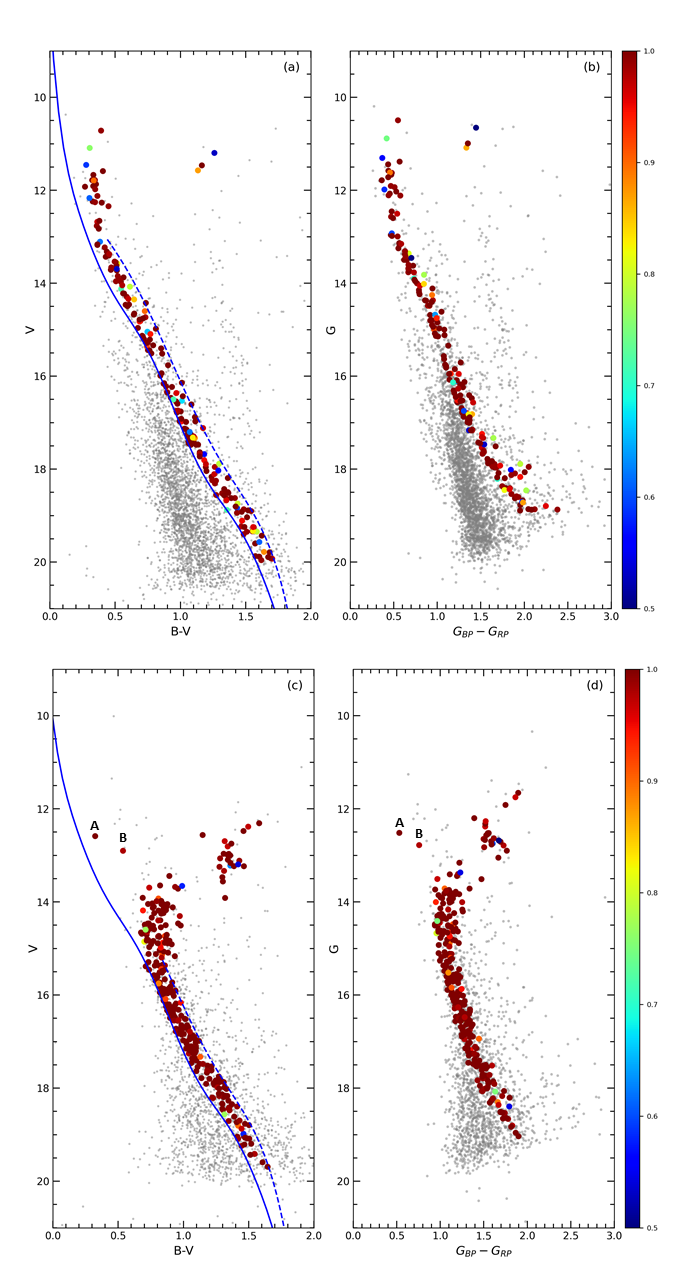}
\caption{$V\times (B-V)$ and $G\times (G_{\rm BP}-G_{\rm RP})$ CMDs of NGC 1664 (a, b) and NGC 6939 (c, d). The blue dot-dashed lines represent the ZAMS \citep{Sung13} including the binary star effect. The membership probabilities of stars that lie within the fitted ZAMS are shown with different colors, these member stars are located within $r_{\rm lim}=8.5$ and $r_{\rm lim}=6.5$ arcmin of the cluster centres calculated for NGC 1664 and NGC 6939, respectively. Grey dots indicate field stars. Letters A and B in panels (c) and (d) indicate possible BSSs of the NGC 6939 open cluster. 
\label{fig:cmds}}
\end {figure*}

% FIGURE 7
\begin{figure}
\centering
\includegraphics[scale=0.8, angle=0]{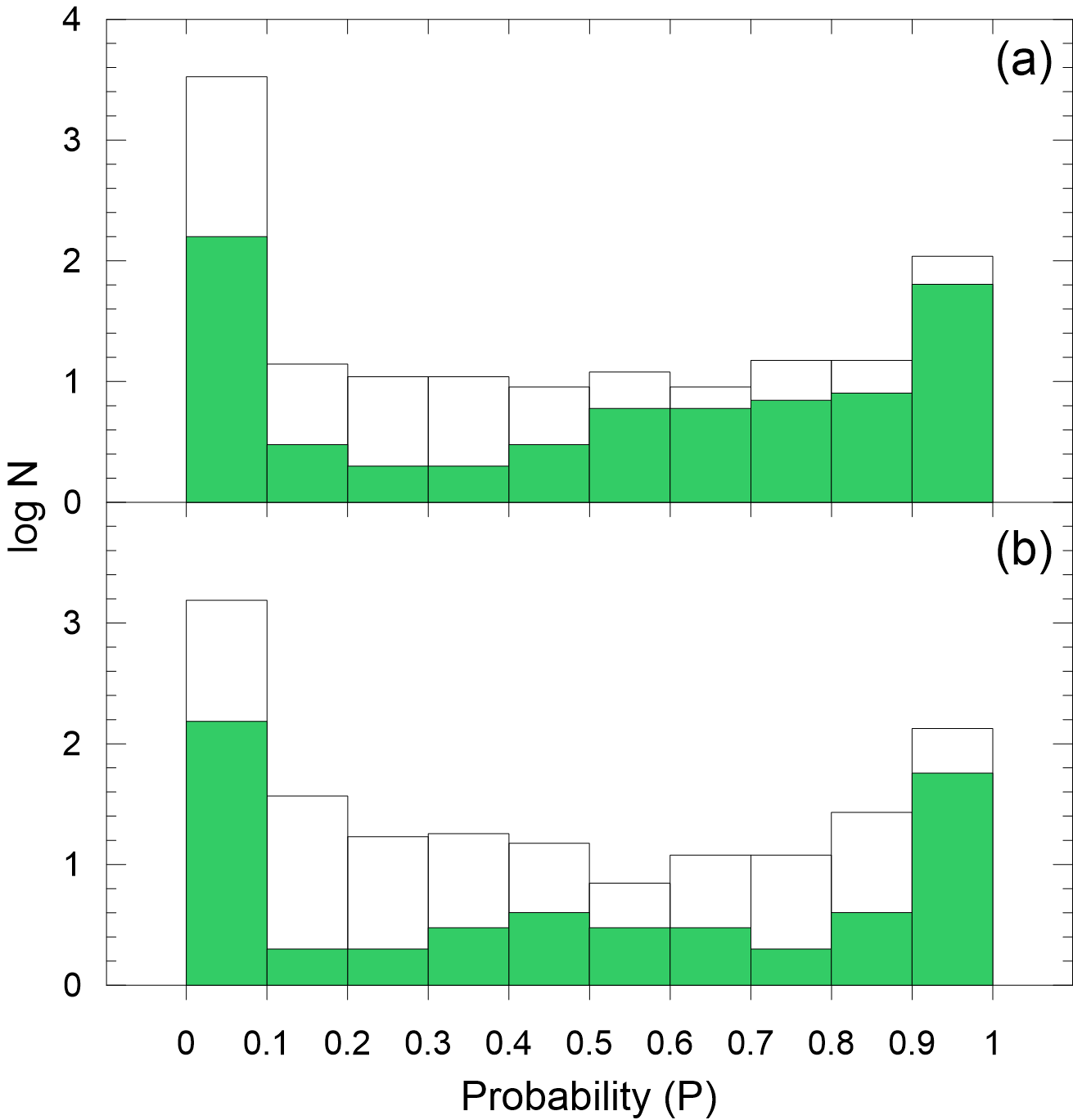}
\caption{Histograms of the membership probabilities versus number of stars for NGC 1664 (a) and NGC 6939 (b). The green colored shading denotes the stars that lie within the main-sequence band and effective cluster radii.
\label{fig:prob_hists} }
\end {figure}

% FIGURE 8
\begin{figure*}
\centering
\includegraphics[scale=.70, angle=0]{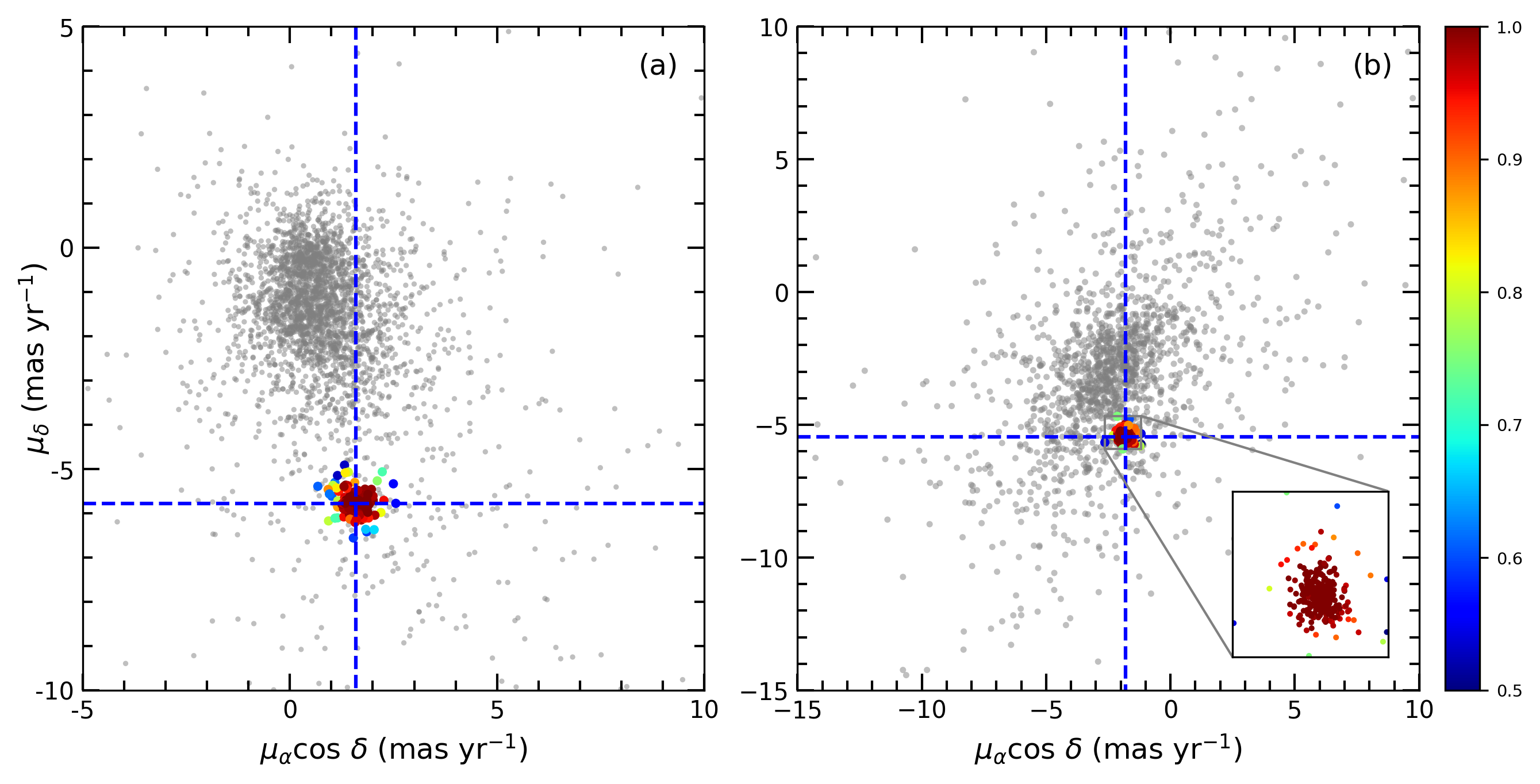}\\
\caption{VPDs of NGC 1664 (a) and NGC 6939 (b). Colored dots identify the membership probabilities of the cluster stars according to color scale shown on the right. The zoomed box in panel (b) represents the region of condensation for the NGC 6939 cluster in the VPD. Dashed lines are the intersection of the mean proper motion values. The color scale shows the membership probabilities of the most likely cluster members.
\label{fig:vpds}} 
\end {figure*}

%----------------------------------------------------------------------------------

\subsection{CMDs and Membership Probabilities of Stars}
\label{section:cmds}
 
Proper motion components of stars are crucial data in order to separate the field stars from the cluster members \citep{Bisht20}. Because cluster stars have a common origin they share similar vectoral movements in the sky. This property makes proper motions a useful tool to eliminate non-member stars from the cluster's main sequence \citep{Yadav13}. Astrometric measurements of {\it Gaia} EDR3 provide proper motion and trigonometric parallaxes of billions of stars, including suspected members of the clusters of interest. Using these data together with the statistical methods developed by various researchers \citep{Balaguer98, Zhao90, Stetson80} allowed us to calculate membership probabilities of stars. 

To obtain the most likely cluster members and to calculate membership probabilities of stars, we used proper motion components ($\mu_{\alpha}\cos \delta$, $\mu_{\delta}$) of {\it Gaia} EDR3 and utilized the method of \citet{Balaguer98}. As a non-parametric method, it arranges the cluster and field stars by trial and error taking into account stellar mean proper motion components as well as their errors. Proper motion measurements of 3,687 and 2,087 stars are given in the current catalog for NGC 1664 and NGC 6939 respectively. Vector point diagrams (VPDs) for these stars are plotted in Fig.~\ref{fig:VPD_all}. It can be clearly seen in the Fig.~\ref{fig:VPD_all}a and Fig.~\ref{fig:VPD_all}c that the cluster stars are distinguished from the field stars. We used stellar counts in circular regions and the maximum density method to determine the centre points of the clusters.   Central proper motion component values were found as ($\mu_{\alpha}\cos\delta$, $\mu_{\delta}$)=(1.64, $-5.85$) mas yr$^{-1}$ for NGC 1664 and as ($\mu_{\alpha}\cos\delta$, $\mu_{\delta}$)=(-1.83, -5.48) mas yr$^{-1}$ for NGC 6939 (see Table~\ref{tab:membership}). 

To determine the radius of the circle we plotted the number of stars versus radial distance of proper motion and fitted the function of stellar density profile for each cluster shown in Fig.~\ref{fig:VPD_all}b and Fig.~\ref{fig:VPD_all}d. Considering the radial distance of proper motion value where the number density is closer to the field star density, we obtained the radius of the circles as 0.8 mas yr$^{-1}$ and 0.5 mas yr$^{-1}$, for NGC 1664 and NGC 6939, respectively. In this way, we obtained 311 and 545 possible member stars for NGC 1664 and NGC 6939, respectively. 

The method of \citet{Balaguer98} defines the membership probability of the i$^{\rm th}$ star as:
\begin{equation}
P_{\rm \mu}(i) = \frac{n_{\rm c} \times \phi_c^{\nu}(i)}{n_{\rm c} \times \phi_{\rm c}^{\nu}(i) + n_{\rm f} \times \phi_{\rm f}^{\nu}(i)}
\end{equation}
where $n_{\rm c}$ and $n_{\rm f}$ represent the normalised number of stars for the regions of cluster and star field ($n_{\rm c} + n_{\rm f} = 1$).  The $\phi_{\rm c}^{\nu}$ and $\phi_{\rm f}^{\nu}$ denote the frequency distribution functions for the cluster and field stars. For the i$^{\rm th}$ star $\phi_{\rm c}^{\nu}$ is described as:\\
\begin{eqnarray}
\phi_{\rm c}^{\nu}(i) =\frac{1}{2\pi\sqrt{{(\sigma_{\rm xc}^2 + \epsilon_{\rm xi}^2)} {(\sigma_{\rm yc}^2 + \epsilon_{\rm yi}^2)}}} \times \exp \left({-\frac{1}{2}\left[\frac{(\mu_{\rm xi} - \mu_{\rm xc})^2}{\sigma_{\rm xc}^2 + \epsilon_{\rm xi}^2 } + \frac{(\mu_{\rm yi} - \mu_{\rm yc})^2}{\sigma_{\rm yc}^2 + \epsilon_{\rm yi}^2}\right]}\right) \\ \nonumber
\end{eqnarray}
\noindent where $\mu_{\rm xi}$ and $\mu_{\rm yi}$ are the proper motion components, while $\epsilon_{\rm xi}$ and $\epsilon_{\rm yi}$ are the relevant proper motion errors for $i^{\rm th}$ star. $\mu_{\rm xc}$ and $\mu_{\rm yc}$ are the proper motion components of the cluster center with dispersion $\sigma_{\rm xc}$ and $\sigma_{\rm yc}$.  $\phi_{\rm f}^{\nu}$ is also defined for the i$^{\rm th}$ star  as:\\
\begin{eqnarray}
\phi_{\rm f}^{\nu}(i) =\frac{1}{2\pi\sqrt{(1-\gamma^2)}\sqrt{{(\sigma_{\rm xf}^2 + \epsilon_{\rm xi}^2 )} {(\sigma_{\rm yf}^2 + \epsilon_{\rm yi}^2 )}}}\\ \nonumber
\times \exp\left({-\frac{1}{2(1-\gamma^2)}}\left[\frac{(\mu_{\rm xi}-\mu_{\rm xf})^2}{\sigma_{\rm xf}^2 + \epsilon_{\rm xi}^2} -\frac{2\gamma(\mu_{\rm xi} - \mu_{\rm xf})(\mu_{\rm yi}-\mu_{\rm yf})} {\sqrt{(\sigma_{\rm xf}^2 + \epsilon_{\rm xi}^2) (\sigma_{\rm yf}^2 + \epsilon_{\rm yi}^2 )}} + \frac{(\mu_{\rm yi}-\mu_{\rm yf})^2}{\sigma_{\rm yf}^2 + \epsilon_{\rm yi}^2}\right]\right)\\ \nonumber
\end{eqnarray}
\noindent where $\mu_{\rm xf}$ and $\mu_{\rm yf}$ represent the field proper motion components with dispersion $\sigma_{\rm xf}$ and $\sigma_{\rm yf}$. The correlation coefficient $\gamma$ is described as\\
\begin{equation}
\gamma = \frac{(\mu_{\rm xi} - \mu_{\rm xf})(\mu_{\rm yi} - \mu_{\rm yf})}{\sigma_{\rm xf}\sigma_{\rm yf}}\\
\end{equation}

% Table 7
\begin{table}
\setlength{\tabcolsep}{10pt}
\renewcommand{\arraystretch}{0.8}
  \centering
  \caption{Mean proper motion values of cluster and field stars on the VPDs.}
    \begin{tabular}{ccc}
    \hline
Parameter         & NGC 1664  &  NGC 6939 \\
    \hline
$n_{\rm c}$                       &  0.08 &  0.26 \\
$n_{\rm f}$                       &  0.92 &  0.74 \\
$\mu_{\rm xc}$ (mas yr$^{-1}$)    & +1.64 & -1.81 \\
$\mu_{\rm yc}$ (mas yr$^{-1}$)    & -5.85 & -5.46 \\
$\sigma_{\rm xc}$ (mas yr$^{-1}$) &  0.25 &  0.14 \\
$\sigma_{\rm yc}$ (mas yr$^{-1}$) &  0.22 &  0.15 \\
$\mu_{\rm xf}$    (mas yr$^{-1}$) & +0.35 & -1.83 \\
$\mu_{\rm yf}$    (mas yr$^{-1}$) & -1.38 & -2.72 \\
$\sigma_{\rm xf}$ (mas yr$^{-1}$) &  1.45 &  1.74 \\
$\sigma_{\rm yf}$ (mas yr$^{-1}$) &  2.15 &  1.50 \\
     \hline
    \end{tabular}%
  \label{tab:membership}%
\end{table}%
\noindent
Determined parameters from the VPDs of two clusters during the membership probability calculations are listed in Table~\ref{tab:membership}.

We selected stars whose membership probabilities $P\geq 0.5$ as the most likely members of the two clusters. This led to the identification of 308 member stars for NGC~1664 and 541 for NGC~6939. Recently \citet{Cantat-Gaudin_Anders20} used {\it Gaia} DR2 data to analyse nearly 2,000 stellar clusters including NGC 1664 and NGC 6939. They identified 299 cluster stars for NGC~1664 and 636 for NGC~6939.  We note that \citet{Cantat-Gaudin_Anders20} utilized the UPMASK method for calculating membership probabilities of the stars and adopted the membership probability $P\geq 0.7$ for the two clusters along with a limiting magnitude $G=18$. Because the methods and data used in the calculation of membership probabilities are different between the current study and \citet{Cantat-Gaudin_Anders20} differences in the member lists could be expected between two studies, but there is a general sense of agreement which is encouraging. For example,  \citet{Cantat-Gaudin_Anders20} used for NGC 6939 a radius of 7.38 arcminutes compared to this study's 6.5, and so would be expected to count more stars.  In the other direction,  \citet{Cantat-Gaudin_Anders20} used a radius of 6.78 arcminutes for NGC 1664, compared to this study's 8.5, and so could be expected to have found less stars.
 
Before starting determination of astrophysical parameters, we also considered `contamination' of the main-sequence by binary stars. We plotted $V\times (B-V)$ color magnitude diagram (CMD) of the stars located in the direction and within the $r_{\rm lim}$ radii of each cluster. We fitted Zero Age Main-Sequence (ZAMS) isochrones \citep{Sung13} to the most likely cluster members ($P\geq0.5$), shifting the fitted ZAMS by 0.75 mag to account for binary star contamination. Attention was paid to the overall fitting, including consideration of not only the main sequence but also the turn-off and giant regions. The $V\times (B-V)$ CMDs with fitted ZAMS are shown as Figs.~\ref{fig:cmds}a and \ref{fig:cmds}c. The distribution of field and member stars with {\it Gaia} EDR3 photometry are shown as $G\times (G_{\rm BP}-G_{\rm RP})$ CMDs in Figs.~\ref{fig:cmds}b and \ref{fig:cmds}d. These restrictions described above in this paragraph, this led to 197 stars identified in NGC~1664 and 279 in NGC~6939.  Figure~\ref{fig:prob_hists} presents histograms of membership probabilities of stars through the regions described for NGC~1664 and NGC~6939. In Figs.~\ref{fig:cmds}c and \ref{fig:cmds}d, the two stars ``A'' ($\alpha=20^{\rm h} 31^{\rm m} 31^{\rm s}.92$, $\delta= +60^{\rm o} 38^{\rm '} 15^{\rm''}.85$) and ``B'' ($\alpha=20^{\rm h} 31^{\rm m} 31^{\rm s}.95$, $\delta= +60^{\rm o} 36^{\rm '} 25^{\rm''}.53$), seen on the blue side of the  main-sequence turn off of the NGC 6939, have estimated membership probabilities of $P=1$ and $P=0.98$, respectively, and appear as blue straggler stars (BSSs) of the cluster. Recently \citet{Jadhav21} and \citet{Rain21} studied BSSs in the open cluster NGC 6939. Their analyses are based on {\it Gaia} DR2 astrometric and photometric data. In their studies, \citet{Jadhav21} and \citet{Rain21} determined that star ``A'' is a BSS and a cluster member. \citet{Jadhav21} calculated the mass of star ``A'' as 3.8 $M/M_{\odot}$. The two studies found no evidence whether star ``B'' is a BSS. The high precision astrometric data of {\it Gaia} EDR3 have increased the sensitivity of stellar membership determinations. This may increase the probability of star ``B'' being a member of the cluster. In this case, star ``B'' appears to be a possible BBS candidate and member of NGC 6939.

We constructed VPDs for each cluster to explore the distribution and separation of the members stars in the proper motion space. Figure~\ref{fig:vpds} demonstrates that the selected member stars were well separated from the field stars for both clusters. We determined mean proper motion components of the adopted member stars and present the intersections of these values on Fig.~\ref{fig:vpds} as dashed lines.
The mean proper motion components were calculated as $\mu_{\alpha}\cos \delta = 1.594 \pm 0.071$ and  $\mu_{\delta} = -5.780 \pm 0.052$ mas yr$^{-1}$ for NGC~1664, and $\mu_{\alpha}\cos \delta = -1.817 \pm 0.039 $ and $\mu_{\delta} = -5.462 \pm 0.039$ mas yr$^{-1}$ for NGC~6939.

%---------------------------------------------------------------
\newpage

\section{Astrophysical Parameters of the Clusters}
In this section we summarize the processes which we performed during the determination of astrophysical parameters of NGC 1664 and NGC 6939 \citep[for detailed descriptions on the methodology see][]{Yontan15, Yontan19, Yontan21, Ak16, Bilir06, Bilir10, Bilir16, Bostanci15, Bostanci18, Banks20, Akbulut2021}. Color excesses and metallicities of the clusters were derived using two color diagrams (TCDs), whereas we obtained distance moduli and ages individually by fitting theoretical models on CMDs. 

%---------------------------------------------------------------

\subsection{Reddening}

\begin{figure*}
\centering
\includegraphics[scale=0.6, angle=0]{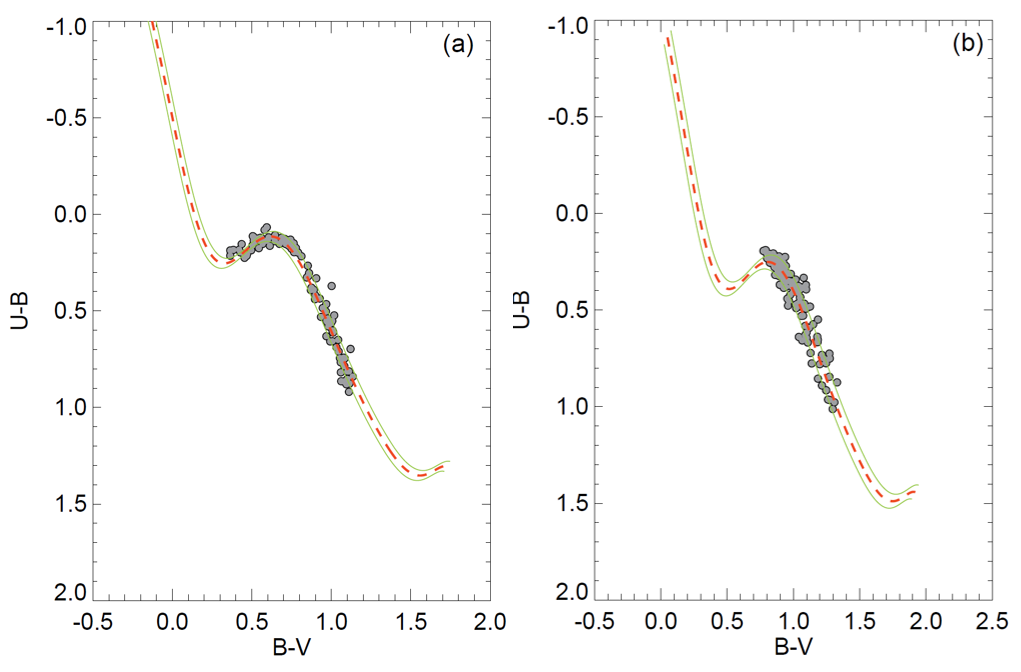}
\caption{Two-color diagrams of the most probable member main-sequence stars in the regions of NGC 1664 (a) and NGC 6939 (b). Red dashed and green solid curves represent the reddened ZAMS given by \citet{Sung13} and $\pm1\sigma$ standard deviations, respectively.
\label{fig:tcds}} 
\end{figure*}

To determine the reddening across each cluster we considered the most likely main-sequence members ($P\geq 0.5$). The $V$ magnitude range of these stars is $13\leq V \leq 18$ for NGC~1664 and $15\leq V \leq 20$ mag for NGC~6939, respectively. Taking into account the distribution of member stars on $(U-B)\times (B-V)$ diagrams, we fitted the solar-metallicity de-reddened ZAMS of \citet{Sung13} to the observational data. We employed $\chi^2$ optimisation with steps of 0.01 mag according to slope of the reddening given by \citet{Garcia88} as ($E(U-B)=0.72 \times E(B-V) + 0.05\times E(B-V)^2$). $E(B-V)$ and $E(U-B)$ values corresponding to the minimum $\chi^2$ were adopted as the best solutions of reddening, being $E(B-V)=0.190\pm 0.018$ mag for NGC 1664 and $E(B-V)=0.380\pm 0.025$ mag for NGC 6939 TCDs with the best result fits are shown as Fig.~\ref{fig:tcds}. The errors of reddening are  estimated as $\pm 1\sigma$ deviations, and are presented with green lines on the same figure.    

%---------------------------------------------------------------

% FIGURE 9

% FIGURE 10
\begin{figure*}
\centering
\includegraphics[scale=0.70, angle=0]{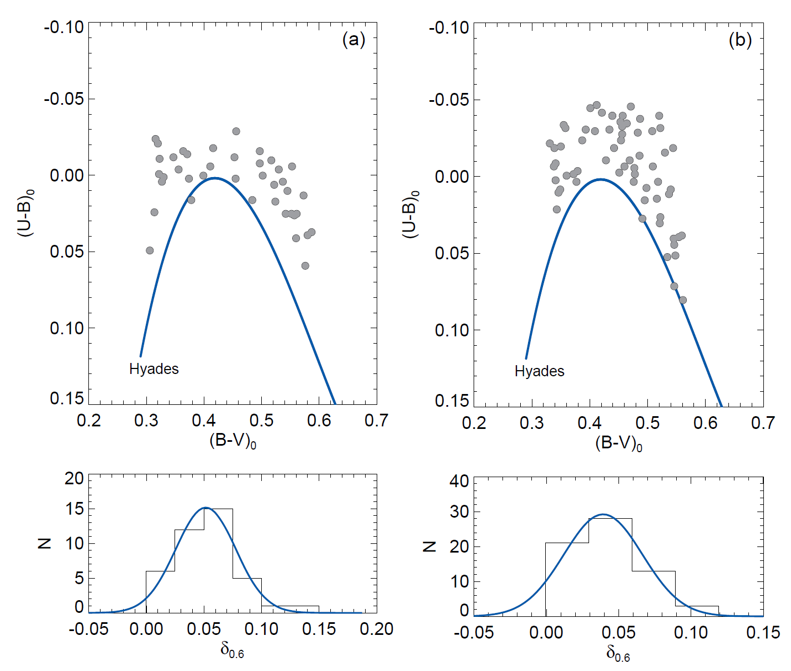}
\caption{$(U-B)_0\times(B-V)_0$ diagrams (upper panels) and the distributions of normalised $\delta_{0.6}$ (lower panels) for NGC 1664 (a) and NGC 6939 (b). The solid blue lines in the upper and lower panels represent the main-sequence of Hyades and Gaussian models which were fitted to the histograms, respectively.
\label{fig:hyades}} 
\end {figure*}

%---------------------------------------------------------------

\subsection{Metallicities}

In order to determine the photometric metallicities we employed the method of \citet{Karaali11}, which is based on F- and G-type main-sequence stars and their UV-excesses. We calculated the intrinsic $(B-V)_0$ and $(U-B)_0$ colors of the most likely members ($P\geq 0.5$) and made a selection of main-sequence stars considering $0.3\leq (B-V)_0\leq0.6$ mag range \citep{Eker20, Eker18} which corresponds to F- and G- type main-sequence stars. Plotting the $(U-B)_0\times(B-V)_0$ TCDs of the selected stars and the Hyades main-sequence, we calculated the difference between the $(U-B)_0$ color indices of the member stars and the Hyades stars, which is defined as the UV-excess ($\delta$). UV-excess is expressed by $\delta =(U-B)_{\rm 0,H}-(U-B)_{\rm 0,S}$ where H and S are the Hyades and cluster stars, respectively, whose intrinsic $(B-V)_0$ colors are the same. We normalised the UV-excess to $(B-V)_0 = 0.6$ mag (i.e. $\delta_{0.6}$) and constructed the histogram according to the $\delta_{0.6}$ values of the stars in order to calculate the mean $\delta_{0.6}$ by fitting Gaussians to the distribution \citep{Karaali03a, Karaali03b}. The photometric metallicity calculation is based on mean $\delta_{0.6}$, which corresponds to the peak of the Gaussian fit, and the relevant equation used in analyses is given as follows \citep{Karaali11}:  
\begin{eqnarray}
{\rm [Fe/H]}=-14.316(1.919)\delta_{0.6}^2-3.557(0.285)\delta_{0.6}+0.105(0.039).
\end{eqnarray}
$(U-B)_0\times(B-V)_0$ diagrams and histograms of $\delta_{0.6}$ are presented in Fig.~\ref{fig:hyades}. The number of most likely member stars which were considered during the metallicity calculations were 40 for NGC 1664 and 65 for NGC 6939. From these stars we determined the photometric metallicity for NGC 1664 as [Fe/H]=$-0.10\pm 0.02$ and for NGC 6939 as [Fe/H]=$-0.06\pm 0.01$ dex. 

To derive ages of the clusters it is required to convert iron abundance to the metallicity of all elements heavier than helium. To do this, we used the analytic equations denoted by Bovy\footnote{https://github.com/jobovy/isodist/blob/master/isodist/Isochrone.py} for {\sc parsec} isochrones \citep{Bressan12}. The equations are expressed as given below: 

\begin{equation}
z_{\rm x}={10^{{\rm [Fe/H]}+\log \left(\frac{z_{\odot}}{1-0.248-2.78\times z_{\odot}}\right)}}
\end{equation}      
and
\begin{equation}
z=\frac{(z_{\rm x}-0.2485\times z_{\rm x})}{(2.78\times z_{\rm x}+1)}.
\end{equation} 
$z$ are the elements heavier than helium, $z_{\rm x}$ is the intermediate operation function, and $z_{\odot}$ is the solar metallicity which was adopted as 0.0152 \citep{Bressan12}. We calculated $z=0.012$ for NGC 1664 and $z=0.013$ for NGC 6939.

\subsection{Distance Moduli and Age Estimation}

\begin{figure*}
\centering
\includegraphics[scale=.9, angle=0]{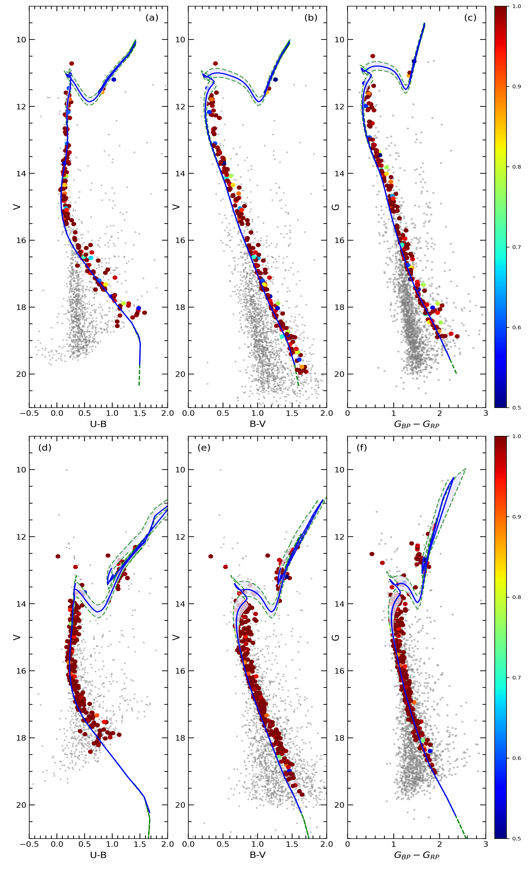}
\caption{CMDs for the NGC 1664 (panels a, b, and c) and NGC 6939 (panels d, e, and f). The differently colored dots represent the membership probabilities according to the color scales shown on the right side of the diagrams. Grey colored dots identify the field stars. The blue lines show the {\sc parsec} isochrones, while the purple shaded areas surrounding these lines are their associated errors. The best fitting isochrone for NGC 1664 corresponds to a 675 Myr age for the cluster, while that for NGC 6939 is 1.5 Gyr.
\label{fig:figure_ten} }
\end {figure*}

The distance moduli and age of the two clusters were determined simultaneously by comparing {\sc parsec} isochrones \citep{Bressan12} to the $V\times (U-B)$, $V\times (B-V)$, and $G\times (G_{\rm BP}-G_{\rm RP})$ CMDs based on likely member stars ($P\geq 0.5$). The selection of {\sc parsec} isochrones was made by taking into account the mass fractions ($z$) calculated for each cluster. We fitted the selected isochrones on CMDs visually placing emphasis on fitting main-sequence, turn-off and giant likely members. For the distance moduli and age estimation for $UBV$ data we shifted the {\sc parsec} isochrones according to the $E(B-V)$ values calculated above. Whereas for the {\it Gaia} EDR3 data we used the equation $E(G_{\rm BP}-G_{\rm RP})= 1.41\times E(B-V)$, where the coefficient was calculated using the equation of \citet{Sun21} who presented selective absorption coefficients. In the error determination of the distance moduli and distances we used the relations of \citet{Carraro17}. We estimated the uncertainty in the derived cluster ages by fitting two more isochrones whose values were good fits to the data sets but at the higher and lower acceptable values compared to the adopted mean age.  The best fit isochrones derive the assumed ages for the clusters, whereas other two closing fitting isochrones, where one is younger and the other is older than the adopted best fit age, were taken as estimating the fitting uncertainties. The best fit with $z=0.012$ gave the distance moduli and age of NGC 1664 as $\mu=11.140 \pm 0.080$ mag and $t=675 \pm 50$ Myr, respectively. For the NGC 6939 the best fit with $z=0.013$ gave these values as $\mu=12.350 \pm 0.109$ mag and $t=1.5 \pm 0.2$ Gyr, respectively. The distances of the clusters corresponding to the estimated distance moduli are also $d_{\rm iso}=1289 \pm 47$ pc for NGC 1664 and $d_{\rm iso}=1716 \pm 87$ pc for NGC 6939 (see Table~\ref{tab:table_one}). The adopted best fit age isochrones and relevant errors are over-plotted on the CMDs which were constructed from $UBV$ and {\it Gaia} EDR3 data and shown in Fig.~\ref{fig:figure_ten}. The discussion of different distance estimations will be interpreted in relation of {\it Gaia} astrometric results in Section~\ref{section:five}.

% FIGURE 11

%---------------------------------------------------------------
%Table 8
\begin{table*}
\setlength{\tabcolsep}{1.6pt}
  \centering
  \caption{\label{tab:seven}
Mean proper motion components and distances ($d_{\rm iso}$, $d_{\rm Gaia~EDR3}$, $d_{\rm BJ21}$) as estimated in this study. The results of Cantat-Gaudin \& Anders (CA20; 2020) and Bailer-Jones (BJ21; 2021) are also listed in the table.}
\begin{tabular}{l|cccc|ccc|c}
\hline
                  & \multicolumn{4}{c|}{This Study} &  \multicolumn{3}{c|}{CA20} & BJ21 \\
\hline
     Cluster      & $d_{\rm iso}$ & $d_{\rm Gaia~EDR3}$ & $\mu_{\alpha}\cos \delta$ & $\mu_{\delta}$ & $d$ & $\mu_{\alpha}\cos \delta$ & $\mu_{\delta}$ & $d_{\rm BJ21}$ \\
                  &    (pc)   & (pc) & (mas~yr$^{-1}$)  & (mas~yr$^{-1}$) & (pc) & (mas~yr$^{-1}$) & (mas~yr$^{-1}$) & (pc) \\
\hline
    NGC 1664 &  1289$\pm$47 &  1350$\pm$81  &+1.594$\pm$0.071 & $-$5.780$\pm$0.052 & $1310^{+198}_{-152}$   & $+1.703\pm0.014$ & $-5.738\pm0.010$ &  1292$\pm$87 \\
    NGC 6939 &  1716$\pm$87 & 1912$\pm$110  & $-$1.817$\pm$0.039 & $-$5.462$\pm$0.039 & $1868^{+429}_{-294}$ & $-1.841\pm0.006$ & $-5.413\pm 0.006$ &  1842$\pm$114 \\
 \hline
    \end{tabular}%
\end{table*}%

%---------------------------------------------------------------

\section{Comparison of Astrometric Results}
\label{section:five}
We also calculated trigonometric distances from {\it Gaia} EDR3 astrometric data \citep{Gaia21} using the most likely member stars of two clusters.

\begin{figure}
\centering
\includegraphics[scale=0.40, angle=0]{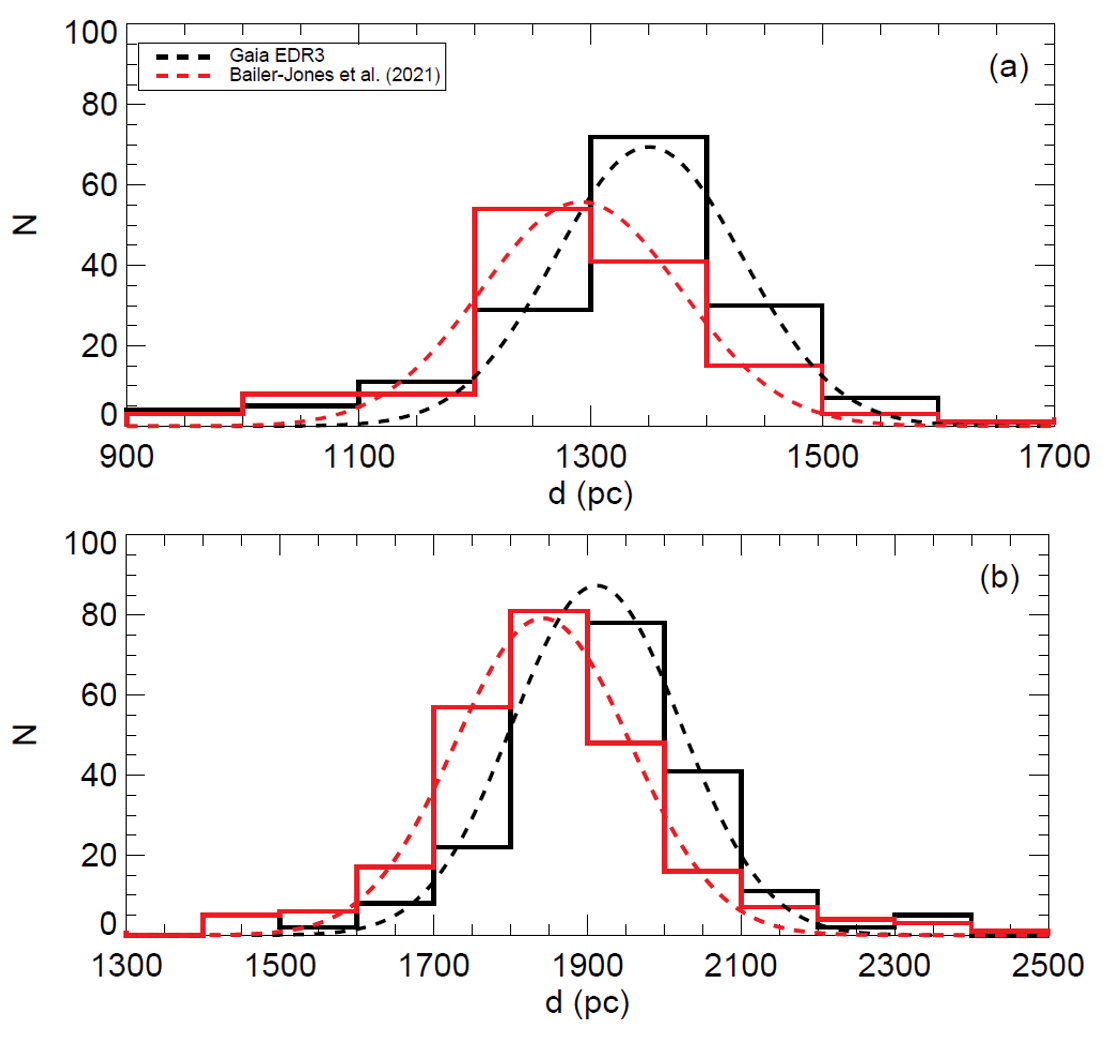}
\caption{The distance histograms of NGC 1664 (a) and NGC 6939 (b). Red and black dashed lines show the best Gaussian fits to the distances Bayesian approach method \citep{Bailer-Jones21} and calculated from parallaxes \citep{Gaia21}.
\label{fig:eleven}}
\end {figure}

Trigonometric distances were determined by converting trigonometric parallaxes into distances, using the linear expression of $d({\rm pc})=1000/\varpi$ (mas) applied to each selected member star of the NGC 1664 and NGC 6939 clusters. In calculating the mean distances,{\it Gaia} stars with a relative parallax error of less than 0.2 were considered. Histograms of obtained distances were constructed and Gaussians fitted to them (Fig.~\ref{fig:eleven}), then taking into account the fitted maxima of Gaussian models we derived mean {\it Gaia} distances ($d_{\rm Gaia~EDR3}$) for each cluster. Uncertainties in distances are one standard deviation. The results of analyses give the {\it Gaia} distances as $d_{\rm Gaia~EDR3}=1350\pm81$ pc for NGC 1664 and $d_{\rm Gaia~EDR3}=1912\pm110$ pc for NGC 6939 (see also Table~\ref{tab:seven}). 

\citet{Bailer-Jones21} (hereafter BJ21) stated in their recent research that transforming trigonometric parallaxes to distances using the linear method does not give accurate measurements. For the precise calculation of trigonometric distances \citet{Bailer-Jones21} developed a geometric approach which considers the probability distributions of stellar distances. We cross-matched the most likely cluster stars with those given in the catalog of BJ21, retrieving their distance estimates, and plotted the histograms of these distances versus number of stars (Fig.~\ref{fig:eleven}). We fitted Gaussian models to each distribution and derived mean distances ($d_{\rm BJ21}$) of the two clusters. Errors for $d_{\rm BJ21}$ distances were taken as the standard deviations of Gaussian fits (see Table~\ref{tab:seven}). As a result we estimated the BJ21 distances as $d_{\rm BJ21}=1292\pm 87$ pc and $d_{\rm BJ21}=1842\pm 114$ pc for NGC 1664 and NGC 6939, respectively. Comparisons of the distance distributions of the two data sets are shown in Fig.~\ref{fig:eleven}.  

We compared distances which were determined via three methods (isochrone fitting, linear expression, and geometric approach); these distances are listed in Table~\ref{tab:seven}. It can be seen that the distances derived from isochrones fitting are compatible with the results of {\it Gaia} and BJ21. It is suggested that trigonometric parallaxes of {\it Gaia} have a small bias given the work by \citep{Groenewegen21}. This could partly clarify why the {\it Gaia} distances are estimated further.     
We also compared the astrometric results ($\mu_{\alpha}\cos \delta$, $\mu_{\delta}$, $d_{\rm iso}$, $d_{\rm Gaia~EDR2}$ and $d_{\rm BJ21}$) of the two clusters with those presented in the study of \citet{Cantat-Gaudin20}. This comparison is given in Table~\ref{tab:seven}. We can conclude that our findings are in agreement with the results of \citet{Cantat-Gaudin20}, being inside the 2$\sigma$ ranges for both clusters.

% FIGURE 12

%---------------------------------------------------------------

\section{Space Velocities and Galactic Orbits of NGC 1664 and NGC 6939}

The galactic orbital parameters of NGC 1664 and NGC 6939 were calculated using the potential functions defined in {\sc galpy}, the galactic dynamics library \citep{Bovy15}\footnote{See also https://galpy.readthedocs.io/en/v1.5.0/}. The calculation assumed an axisymmetric potential for the Milky Way galaxy, following {\sc MWPotential2014} \citep{Bovy15}. The galactocentric distance and a circular velocity of the Sun were assumed as $R_{\rm GC}=8$ kpc and $V_{\rm rot}=220$ km s$^{-1}$, respectively \citep{Bovy15, Bovy12}. The distance from the galactic plane of the Sun was adopted as 27$\pm$4 pc \citep{Chen00}.

We adopted that the Galaxy is well represented by the {\sc MWPotential2014} code, which is composed of the bulge, disc, and halo potentials of the Milky Way. The bulge component is represented as a spherical power law density profile by \cite{Bovy15}: 

\begin{equation}
\rho (r) = A \left( \frac{r_{\rm 1}}{r} \right) ^{\alpha} \exp \left[-\left(\frac{r}{r_{\rm c}}\right)^2 \right] \label{eq:rho}
\end{equation} 
where $r_{\rm 1}$ is the present reference radius and $r_{\rm c}$ the cut-off radius. $A$ is the amplitude that is applied to the potential in mass density units and $\alpha$ presents the inner power. We used the potential proposed by \citet{Miyamoto75} for the galactic disc component:

\begin{equation}
\Phi_{\rm disc} (R_{\rm GC}, Z) = - \frac{G M_{\rm d}}{\sqrt{R_{\rm GC}^2 + \left(a_{\rm d} + \sqrt{Z^2 + b_{\rm d}^2 } \right)^2}} \label{eq:disc}
\end{equation}
$R_{\rm GC}$ is the distance from the galactic centre, $Z$ is the vertical distance from the galactic plane, $G$ is the universal gravitational constant, $M_{\rm d}$ is the mass of the galactic disc, $a_{\rm d}$ is the scale length of the disc, and $b_{\rm d}$ is the scale height of the disc. 

The potential for the halo component was obtained from \citet{Navarro96}:

\begin{equation}
\Phi _{\rm halo} (r) = - \frac{G M_{\rm s}}{R_{\rm GC}} \ln \left(1+\frac{R_{\rm GC}}{r_{\rm s}}\right) \label{eq:halo}
\end{equation} 
where $M_{\rm s}$ if the mass of the dark matter halo of the Milky Way and $r_{\rm s}$ is its radius.

The {\sc MWPotential2014} code was used to determine the space velocity components and galactic orbital parameters for both OCs. The equatorial coordinates, proper motion components, distance, and radial velocity data required for the calculations of the space velocities and galactic orbit parameters of the clusters are listed in Table~\ref{tab:input_parameters2}. Kinematic and dynamic calculations were analyzed with 1 Myr steps over a 3.5 Gyr integration time.

% TABLE 9
\begin{table*}
\setlength{\tabcolsep}{0.5pt}
\renewcommand{\arraystretch}{0.8}
  \centering
  \caption{\label{tab:input_parameters2}
  The input and obtained output parameters for the orbit integration of NGC 1664 and NGC 6939.}
    \begin{tabular}{lcccccccc}
\hline
\multicolumn{9}{c}{Input Parameters}\\
\hline
\hline
Cluster  &   $\alpha$ (J2000) & $\delta$ (J2000)	& $\mu_{\alpha}\cos\delta$ & 	$\mu_{\delta}$	 & 	$d$	& $V_{\rm r}$ &  & \\
         &  (hh:mm:ss) & (dd:mm:ss) &       (mas yr$^{-1}$)    &    (mas yr$^{-1}$)   & (pc) & (km s$^{-1}$) &  &  \\
\hline
NGC 1664 & 	04:51:06   & 43:40:30	&  +1.594$\pm$0.071         & -5.780$\pm$0.052 & 1289$\pm$47 &   +8.96$\pm$0.24&  & \\	
NGC 6939 & 	20:31:30   & 60:39:42	& -1.817$\pm$0.039          & -5.462$\pm$0.039 & 1716$\pm$87 & -18.91$\pm$0.11& & \\		
\hline
\multicolumn{9}{c}{Output Parameters}\\
\hline
\hline
Cluster  &  $R_{\rm a}$ & $R_{\rm p}$ & $e$ & $Z_{\rm max}$	& $U$	        & $V$           & $W$           & $T$\\
         &  (kpc) & (kpc) &     & (pc)     & (km s$^{-1}$) & (km s$^{-1}$) & (km s$^{-1}$) & (Myr)\\
\hline
NGC 1664 & 	9.47$\pm$0.05   & 7.81$\pm$0.05	&  0.096$\pm$0.001 & 140$\pm$1 & -19.21$\pm$0.61 & -29.90$\pm$1.09 &  -15.56$\pm$0.01 & 244$\pm$1 \\
NGC 6939 &  9.41$\pm$0.11   & 8.23$\pm$0.04	&  0.067$\pm$0.003 & 460$\pm$29 & +46.39$\pm$1.80 & -11.01$\pm$0.45 & -16.63$\pm$0.70 & 248$\pm$2 \\		
\hline
    \end{tabular}
\end{table*}

% FIGURE 13
\begin{figure*}
\centering
\includegraphics[width=\textwidth]{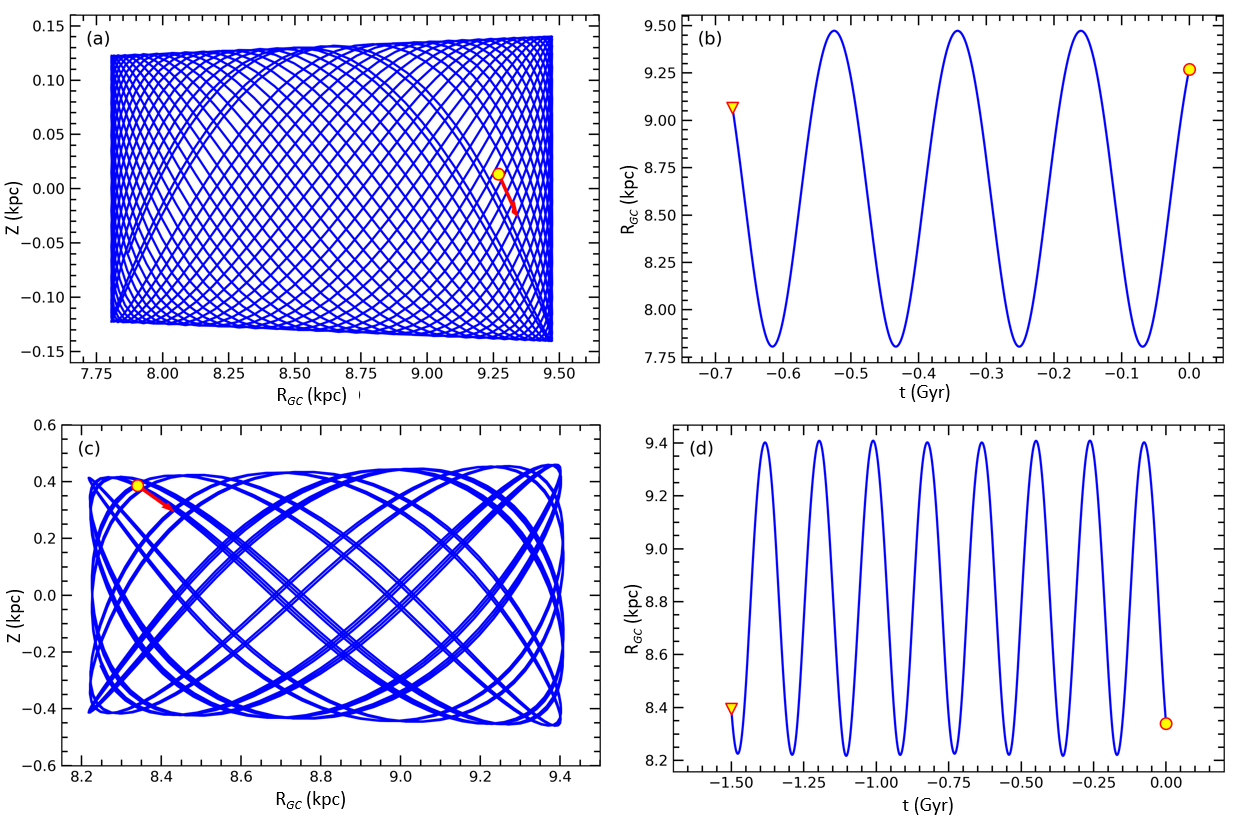}
\caption{\label{fig:galactic_orbits}
The galactic orbits and birth radii of NGC 1664 (a,b) and NGC 6939 (c,d) in the $Z \times R_{\rm GC}$ and $R_{\rm GC} \times t$  planes. The filled yellow circles and triangles show the present day and birth positions, respectively. Red arrows are the motion vectors of OCs.} 
\end {figure*}

The proper motion components and distances of the OCs were calculated in this study and given in Section~\ref{section:cmds}. In order to calculate the mean radial velocities of the two clusters, stars with radial velocity data available in the {\it Gaia} EDR3 catalog and cluster membership $P\geq 0.5$ were selected. Radial velocity data of three stars for NGC 1664 and 22 stars for NGC 6939 were identified in the {\it Gaia} EDR3 catalogue. Using these data we calculated the cluster radial velocities by taking the weighted average of the radial velocities and their uncertainties \citep[for equations see also][]{Soubiran18}. As a result of the analysis, the mean radial velocities were estimated for NGC 1664 as +8.96$\pm$0.24 and for NGC 6939 as $-18.91 \pm 0.11$ km s$^{-1}$. Literature mean radial velocity estimates for NGC 1664 are 6.38$\pm$0.45 \citep{Soubiran18}, +6.38$\pm$0.28 \citep{Carrera19}, 6.37$\pm$0.67 \citep{Tarricq21} and 7.83$\pm$0.03 km s$^{-1}$ \citep{Hao21}. These results are compatible within 1-2 km s$^{-1}$ with the value found for NGC 1664 in this study. A similar situation is also valid for the NGC 6939 cluster, with values given by various researchers being $-21.00 \pm 0.20$ \citep{Warren09}, $-18.99 \pm 0.13$ \citep{Soubiran18}, $-18.60 \pm 0.09$ \citep{Tarricq21} and $-18.77 \pm 0.08$ km s$^{-1}$ \citep{Dias21}. These are compatible with the radial velocities we obtained.

The derived galactic orbit parameters of the two clusters are also listed in Table~\ref{tab:input_parameters}. The parameters estimated from this orbital integration (including uncertainties in distances, proper motions and radial velocities) are apogalactic ($R_{\rm a}$) and perigalactic ($R_{\rm p}$) distances, eccentricity ($e$), the maximum vertical distance from galactic plane ($Z_{\rm max}$), galactic space velocity components ($U$, $V$, $W$), and orbital period ($T$) of each cluster. 

The space velocity components of the NGC 1664 and NGC 6939 were calculated as $(U, V, W)=(-19.21\pm0.61, -29.90\pm1.09, -15.56\pm0.01$) and $(U, V, W)$=($+46.39\pm1.80$, $-11.01\pm0.45$, $-16.63\pm0.70$) km s$^{-1}$, respectively (see also Table~\ref{tab:input_parameters}). As for the local standard of rest (LSR) correction we considered the values of ($8.83\pm0.24$, $14.19\pm0.34$, $6.57\pm0.21$) km s$^{-1}$ which were given by \citet{Coskunoglu11}. We applied LSR corrections to our findings and derived LSR corrected space velocity components as $(U, V, W)_{\rm LSR}$ = ($-10.38\pm0.66$, $-15.71\pm1.14$, $-8.99\pm0.11$) for NGC 1664 and $(U, V, W)_{\rm LSR}$ = ($55.22\pm1.82$, $3.18\pm0.56$, $-10.6\pm0.73$) km s$^{-1}$ for NGC 6939. \citet{Soubiran18} studied these two clusters with {\it Gaia} DR2 \citep{Gaia18} astrometric data finding that the space velocity components of NGC 1664 are $(U, V, W)$=($-16.70\pm0.61$, $-30.49\pm0.23$, $-14.64\pm0.11$) and for NGC 6939 are $(U, V, W)$=($50.13\pm0.19$, $-10.55\pm0.13$, $-17.36\pm0.08$) km s$^{-1}$. It can be seen that the results derived in this study (see Table~\ref{tab:input_parameters}) are in good agreement with the values of \citet{Soubiran18}. The total space velocities are 20.87$\pm$1.32 km s$^{-1}$ for NGC 1664 and 56.32$\pm$2.04 km s$^{-1}$ for NGC 6939. Considering these results, they are compatible with the velocity given by \citet{Leggett92} for young thin-disc stars.

The galactic orbits for the two clusters on the $Z \times R_{\rm GC}$ and $R_{\rm GC} \times t $ planes are shown as panes in Fig.~\ref{fig:galactic_orbits}. Figures~\ref{fig:galactic_orbits}a and \ref{fig:galactic_orbits}c represent `side views' of the cluster motions as functions of distance from the galactic center and the galactic plane. Figures~\ref{fig:galactic_orbits}b and \ref{fig:galactic_orbits}d represent the distances from the galactic plane as a function of time for the two clusters. In Figures \ref{fig:galactic_orbits}b and \ref{fig:galactic_orbits}d we marked the birth and present-day locations for both clusters using yellow filled triangles and circles. Considering the perigalactic and apogalactic distances of the clusters, it can be seen that orbit of the NGC 6939 is completely outside the solar circle (Fig.~\ref{fig:galactic_orbits}c), while NGC 1664 enters the solar circle during its orbit (Fig.~\ref{fig:galactic_orbits}a). The orbits of the two clusters differ from circular. Their eccentricities do not exceed the value of 0.1. The vertical distances from the galactic plane reach at maximum at $Z_{\rm max}=140\pm1$ pc with an orbital period $T=244\pm1$ Myr for NGC 1664 and $Z_{\rm max}=460\pm29$ pc and $T=248\pm2$ Myr for NGC 6939. Considering the relative errors in the galactic model parameters, it can be seen that the errors of most parameters vary between 1\% and 2\%. Only the relative error of the $Z_{\rm max}$ parameter of the NGC 6939 open cluster is about of the order of 6\%. These results show that the orbital parameters calculated for the clusters are precise. In conclusion, it shows that both of the clusters belong to thin disc of Galaxy.

The birthplaces of the clusters were investigated by running the cluster ages calculated in this study back in time using the {\sc galpy} program. In orbital calculations, the birth radii of the OCs NGC 1664 and NGC 6939 were calculated as 9.07 kpc and 8.34 kpc, respectively, assuming that the clusters do not interact with other objects during their lifetimes (see Figs. \ref{fig:galactic_orbits}b and \ref{fig:galactic_orbits}d). These results show that the clusters were born in the metal-poor region outside the solar circle.

%---------------------------------------------------------------

\section{Dynamical Study of the Clusters}

\subsection{Luminosity Functions}
\begin{figure}
\centering
\includegraphics[scale=1.5, angle=0]{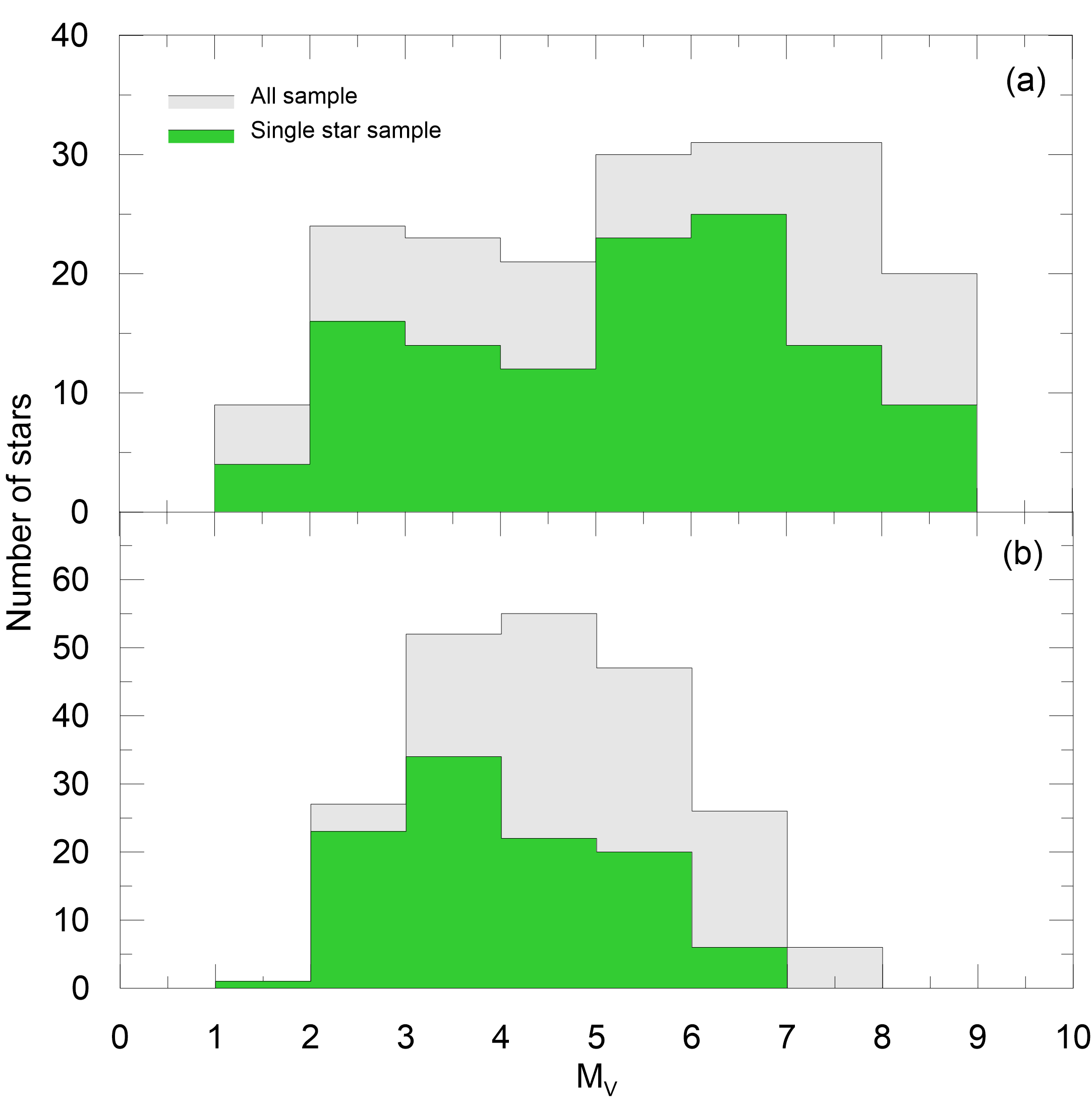}
\caption{\label{fig:luminosity_functions}
The luminosity functions of NGC 1664 (a) and NGC 6939 (b). Grey color shows all sample while green color represents most probable single main-sequence stars}. 
\end {figure}

The luminosity function (LF) is defined as a distribution of stars with magnitude.  To estimate the cluster LFs, we selected the main-sequence stars within the range $12.5 \leq V \leq 20$ mag and which are located inside a 8.5 arcmin limiting radius from the previously estimated centre of the NGC 1664 cluster, and $15 \leq V \leq 20$ mag stars located inside a 6.5 arcmin radius from the centre of the NGC 6939 cluster. We calculated absolute magnitudes for these selected stars using the equation $M_{\rm V} = V-5\times \log d +3.1\times E(B-V)$ where $V$, $d$ and $E(B-V)$ are apparent magnitude, distance, and color excess of each cluster, respectively, as derived earlier in this study (see Table~\ref{tab:table_one}). These calculations resulted in the absolute magnitude ranges being $1< M_{\rm V}< 9$ and $2< M_{\rm V}< 8$ mag for NGC 1664 and NGC 6939, respectively. The resulting LF histograms for both clusters are shown as Fig.~\ref{fig:luminosity_functions}. This figure show that the LFs continue to increase (in stars) up to $M_{\rm V}=8$ mag and $M_{\rm V}=5$ mag for NGC 1664 and NGC 6939, respectively, and begin to drop beyond these limits. In order to analyze the effects on the luminosity and present day mass functions caused by the presence of binary stars in the clusters, the main-sequence band has been reduced from 0.75 to 0.35 magnitude. As a result of this analysis, the numbers of possible single main-sequence stars in the open clusters NGC 1664 and NGC 6939 were determined as 112 and 106, respectively. The distributions of these stars are shown as green-shaded histograms in Fig.~\ref{fig:luminosity_functions}.

% FIGURE 14

%---------------------------------------------------------------

\subsection{Present Day Mass Functions}

To derive Present Day Mass Functions (PDMFs), we used the {\sc parsec} isochrones that best represent the age and metal abundance ($z$) of the clusters to converted the LFs described above to PDMFs. Theoretical models provided by the {\sc parsec} synthetic stellar library \citep{Bressan12} were used to convert the $V$-band absolute magnitudes to masses of theoretical main-sequence stars via a high degree of a polynomial equation in this study. Then we transformed the observational absolute $V$ band magnitudes to masses by considering the absolute magnitude-mass relation. The number, mass range, and mean mass of main-sequence stars so transformed are 189, $0.6\leq M/ M_{\odot}\leq 2$, and 1.13 $M/M_{\rm \odot}$ for NGC 1664, and 213, $0.8\leq M/ M_{\odot}\leq 1.6$, and 1.10 $M/M_{\rm \odot}$ for NGC 6939.

\begin{figure}
\centering
\includegraphics[scale=1.5, angle=0]{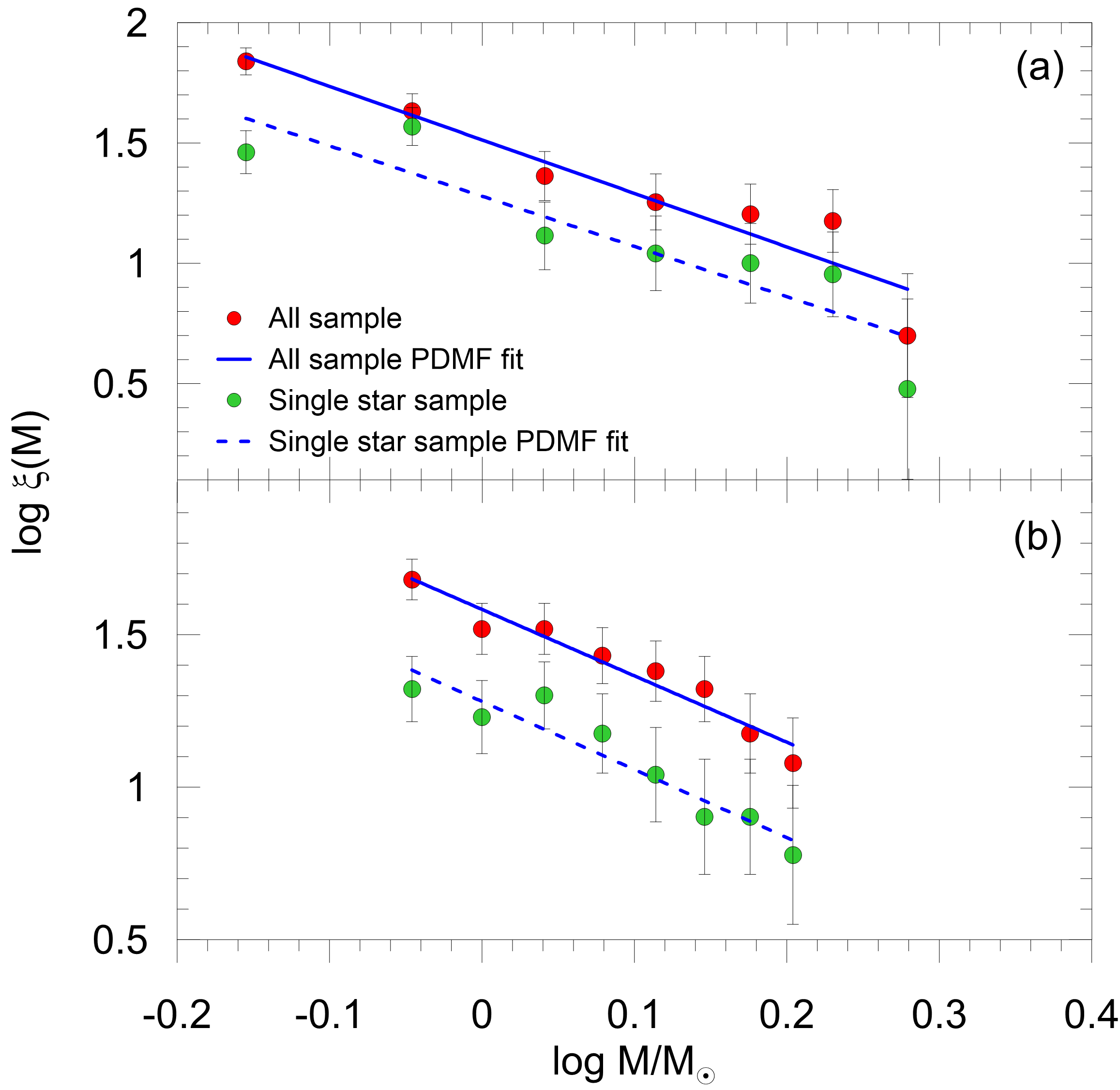}
\caption{\label{fig:mass_functions}
Present day mass functions of NGC 1664 (a) and NGC 6939 (b) derived from all sample (red circle) and probable single star sample (green circle) groups. The mass functions of the OCs are shown with the blue lines.} 
\end {figure}

Mass function slopes of the two clusters were derived using the equation $\log(dN/dM)=-(1+\Gamma)\times \log M + {\rm constant}$, where $dN$ represents the number of stars in a mass interval $dM$ with central mass $M$ and $\Gamma$ the slope of the PDMF. Slopes of the PDMFs were determined as $\Gamma=-1.22\pm0.33$ for NGC 1664 and as $\Gamma=-1.18\pm 0.21$ for NGC 6939. Moreover, in order to analyze the effects of binary stars on present day mass functions, the main-sequence bands were narrowed from 0.75 to 0.35 magnitudes and PDMF values were calculated from possible single main-sequence stars in the clusters. Distributions of PDMF of possible single main-sequence stars that are members of the NGC 1664 and NGC 6939 have been carried out and by applying linear fits to the mass distributions, the slopes of the mass functions have been obtained, respectively, as $\Gamma= -1.09\pm0.45$ and $\Gamma = -1.23\pm0.30$. In general, the slopes of the mass functions calculated from the main-sequence (all sample) and possible single main-sequence stars in both clusters are close to each other, but also very close to \citet{Salpeter55}'s value of $-1.35$. The best fits of PDMFs are shown in Fig.~\ref{fig:mass_functions}.

% FIGURE 15

%---------------------------------------------------------------
\subsection{The Dynamical State of Mass Segregation}

The time scale at which all traces of the initial conditions with which a cluster is born are lost is represented by the relaxation time. The relaxation time ($T_{\rm E}$) is the characteristic time-scale for a cluster to reach the level of energy equipartition. The relaxation time is given by \citet{Spitzer71} as:

\begin{equation}
T_{\rm E} = \frac{8.9 \times 10^{5} N^{1/2} R_{\rm h}^{3/2}}{\langle m\rangle^{1/2}\log(0.4N)}
\end{equation} 
where $N$ is the number of stars, $R_{\rm h}$ is the half-mass radius of the cluster, and $\langle m \rangle$ is the mean mass of the stars in each cluster. The values of $R_{\rm h}$ were taken as 1.67 pc for NGC 1664 and as 1.83 pc for NGC 6939, as calculated earlier in this study. Consequently, the dynamical relaxation times of NGC 1664 and NGC 6939 were derived as $T_{\rm E}=13.48$ Myr and $T_{\rm E}=15.93$ Myr, respectively. These results imply that the age of NGC 1664 is 50 times its relaxation time, and 95 times for NGC 6939. Therefore, we can conclude that the both of clusters are dynamically relaxed.

\begin{figure}
\centering
\includegraphics[scale=0.70, angle=0]{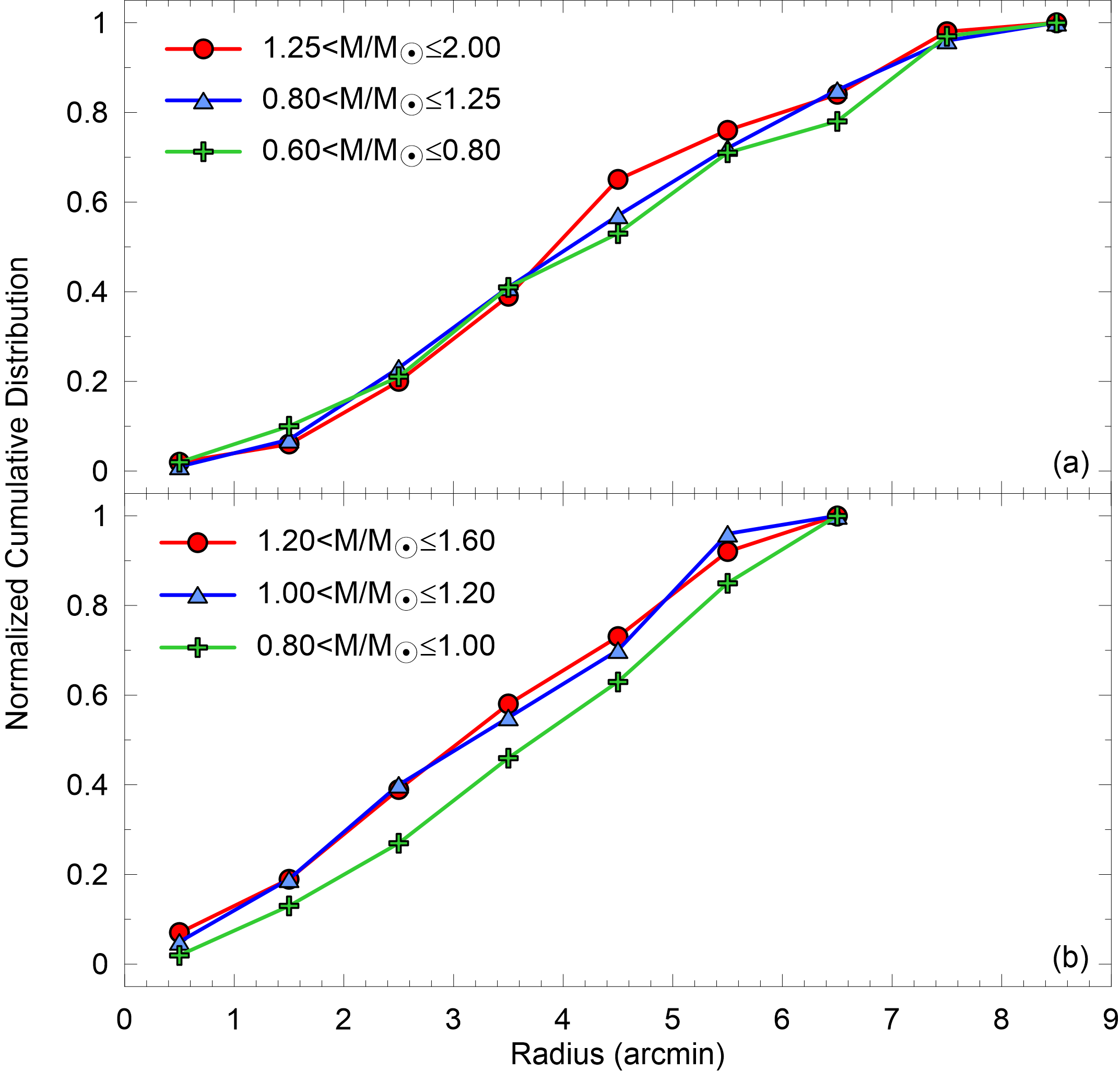}
\caption{\label{fig:radial_distributions}
The cumulative radial distribution of stars in different mass ranges for NGC 1664 (a) and NGC 6939 (b).}
\end {figure}

To investigate whether the clusters have internal mass segregation, we split the cluster star masses into three ranges. These divisions are 1.25 $<M/M_{\odot}\leq$ 2.00 (high mass), 0.80 $<M/M_{\odot}\leq$ 1.25 (intermediate mass), and 0.60 $<M/M_{\odot}\leq$ 0.80 (low mass) for NGC 1664, and 1.20 $<M/M_{\odot}\leq$ 1.60 (high mass), 1.00 $<M/M_{\odot}\leq$ 1.20 (intermediate mass), and 0.80 $<M/M_{\odot}\leq$ 1.00 (low mass) for NGC 6939. The normalized cumulative distributions versus radius from the cluster center of stars in different mass ranges are shown in Fig.~\ref{fig:radial_distributions}.  

It can be seen in Fig.~\ref{fig:radial_distributions}a that increase trend of the stars within the three mass groups is similar from the center to outward of NGC 1664. Considering NGC 6939, cluster stars within high and intermediate masses are concentrated into the cluster center, while the cumulative number of the stars within the low-mass ranges increase outward from the cluster center (Fig.~\ref{fig:radial_distributions}b). In order to understand the statistical significance of mass segregation in both clusters, we performed the Kolmogorov-Smirnov test for the three mass ranges given above. The level of confidence for mass segregation effects are 72\% and 89\% for NGC 1664 and NGC 6939, respectively. 

% FIGURE 16

%----------------------------------------------------------------------------------------------------------

\section{Summary and Conclusion}

We performed a study based on CCD {\it UBV} and {\it Gaia} EDR3 photometric data, as well as {\it Gaia} EDR3 astrometry, of the open clusters NGC 1664 and NGC 6939. We summarise the results of the analyses as follows:

1) We analysed the spatial structure of the two clusters by fitting the RDP of \citet{King62} and obtained limiting radii as $r_{\rm lim}=8'.5$ (3.28 pc) and $r_{\rm lim}=6'.5$ (3.25 pc) for NGC 1664 and NGC 6939, respectively. These values are the point where stellar densities merge with the background stellar density. In the following analyses we considered only the stars inside these limiting radii as potential cluster members.

2) To eliminate background stellar contamination and identify the most likely cluster members, we calculated membership probabilities of stars in the direction of the two clusters. Calculations were based on the stars' proper motion components and their errors from the {\it Gaia} EDR3. We used the probability $P\geq0.5$ as the condition for cluster membership. To take into account the binary star contamination in the cluster main-sequences, we fitted the de-reddened ZAMS of \citet{Sung13} to $V\times (B-V)$ CMDs with a shift of 0.75 mag in the $V$ band. Hence, we assumed the stars within defined effective radii, close to the best-fitting ZAMS curves, and with the membership probability $P\geq0.5$ as the most likely member stars for the two clusters.  

3) We determined reddening and metallicities from CCD {\it UBV} TCDs for the most likely member stars separately to estimate cluster distance modulus and age. The reddening and photometric metallicity for NGC 1664 are $E(B-V)=0.190\pm 0.018$ mag and [Fe/H]=$-0.10\pm 0.02$ dex, respectively. The corresponding values for NGC 6939 are $E(B-V)=0.380\pm 0.025$ mag and [Fe/H]=$-0.06\pm 0.01$ dex.

4) Keeping as constants the reddening and metallicities, we obtained distance moduli, and hence distances and ages of the two clusters by fitting {\sc parsec} isochrones on the {\it UBV} and {\it Gaia} EDR3 photometric CMDs. We found the distance modulus for NGC 1664 to be $\mu_{\rm V}=11.140 \pm 0.080$ mag, which corresponds to the distance being $d=1289 \pm 47$ pc together with an age $t=675 \pm 50$ Myr. For NGC 6939 we estimated $\mu_{\rm V}=12.350\pm 0.109$ mag, which corresponds to a distance $d=1716 \pm 87$ pc and $t=1.5 \pm 0.2$ Gyr. These results are in agreement with those given by different authors.

5) We calculated the mean proper motion components for NGC 1664 as ($\mu_{\alpha}\cos \delta, \mu_{\delta}) = (1.594\pm 0.071, -5.780\pm 0.052$) mas yr$^{-1}$, and for NGC 6939 as $(\mu_{\alpha}\cos \delta, \mu_{\delta})$ = $(-1.817\pm 0.039, -5.462\pm 0.039$) mas yr$^{-1}$. 

6) We determined photometric-based distances by fitting isochrones on CMDs ($d_{\rm iso}$), astrometric data based distance using {\it Gaia} EDR3 trigonometric parallaxes ($d_{\rm Gaia~EDR3}$), and the distance of cluster members that were presented by \citet{Bailer-Jones21} $d_{\rm BJ21}$. We showed that the results are in a good agreement across these methods, and that photometric-based distances are reliable if the cluster member selection is made carefully.

7) Mean radial velocities of the two clusters were derived from the most likely member stars ($P\geq0.5$) that have radial velocities given in the {\it Gaia} EDR3. We found these velocities as $V_{\rm r}=+8.96\pm 0.24$ km s$^{-1}$ for NGC 1664 and $V_{\rm r}=-18.91\pm 0.11$ km s$^{-1}$ for NGC 6939 which are compatible with the results given by different researchers.

8) Space velocities and galactic orbital parameters show that both of the clusters belong to the galactic disc. NGC 1664 orbits in the solar circle while NGC 6939 orbits completely outside it.

9) The birth radii of NGC 1664 and NGC 6939 are 9.07 kpc and 8.34 kpc from the galactic centre, respectively, which indicate that two clusters were born outside the solar circle in a metal-poor region.

10) Present day mass functions were estimated as $\Gamma=-1.22\pm 0.33$ for NGC 1664 and $\Gamma=-1.18 \pm 0.21$ for NGC 6939. These mass function slopes are compatible with the value of \citet{Salpeter55}.

11) We calculated the relaxation times of both clusters and concluded that two clusters are dynamically relaxed. We found no statistical evidence for mass segregation in either cluster.

\software{IRAF \citep{Tody86, Tody93}, PyRAF \citep{Science12}, SExtractor \citep{Bertin96}, Astrometry.net \citep{Lang10}, GALPY \citep{Bovy15}, MWPotential2014 \citep{Bovy15}.}

%---------------------------------------------------------------

\acknowledgments

We thank the anonymous referee for his/her insightful and constructive suggestions, which significantly improved the paper. This study has been supported in part by the Scientific and Technological Research Council (T\"UB\.ITAK) 120F265. We thank T\"UB\.ITAK for partial support towards using the T100 telescope via project 18CT100-1396. We also thank the on-duty observers and members of the technical staff at the T\"UB\.ITAK National Observatory for their support before and during the observations. This research has made use of the WEBDA database, operated at the Department of Theoretical Physics and Astrophysics of the Masaryk University. This research made use of VizieR and Simbad databases at CDS, Strasbourg, France. We made use of data from the European Space Agency (ESA) mission \emph{Gaia}\footnote{https://www.cosmos.esa.int/gaia}, processed by the \emph{Gaia} Data Processing and Analysis Consortium (DPAC)\footnote{https://www.cosmos.esa.int/web/gaia/dpac/consortium}. Funding for DPAC has been provided by national institutions, in particular the institutions participating in the \emph{Gaia} Multilateral Agreement. IRAF was distributed by the National Optical Astronomy Observatory, which was operated by the Association of Universities for Research in Astronomy (AURA) under a cooperative agreement with the National Science Foundation. PyRAF is a product of the Space Telescope Science Institute, which is operated by AURA for NASA. We thank the University of Queensland for collaboration software.

%%%%%%%%%%%%%%%%%%%% REFERENCES %%%%%%%%%%%%%%%%%%


\begin{thebibliography}{99}

\bibitem[Ak et al.(2016)]{Ak16}
Ak, T., Bostanc{\i}, Z. F., Yontan, T., et al., 2016, \apss, 361, 126 

\bibitem[Akbulut et al.(2021)]{Akbulut2021} 
Akbulut, B., Ak, S., Yontan, T., et al., 2021, \apss, 366, 68

\bibitem[Andreuzzi et al.(2004)]{Andreuzzi04}
Andreuzzi, G., Bragaglia, A., Tosi, M., Marconi, G., 2004, \mnras, 348, 297

\bibitem[Bailer-Jones et al.(2021)]{Bailer-Jones21}
Bailer-Jones, C. A. L., Rybizki, J., Fouesneau, M., Mantelet, G., Andrae, R., 2021, \aj, 161, 147

\bibitem[Balaguer-Nunez et al.(1998)]{Balaguer98}
Balaguer-Nunez, L., Tian, K. P., Zhao, J. L., 1998, Astron. Astrophys. Suppl. Ser., 133, 387

\bibitem[Banks et al.(2020)]{Banks20} 
Banks, T., Yontan, T., Bilir, S., Canbay, R., 2020, JApA, 41, 6

\bibitem[Becker \& Fenkart(1971)]{Becker71} 
Becker, W., Fenkart, R., 1971, Astron. Astrophys. Suppl. Ser., 4, 241

\bibitem[Bertin \& Arnouts(1996)]{Bertin96}
Bertin, E., Arnouts, S., 1996, Astron. Astrophys. Suppl. Ser., 117, 393 

\bibitem[Bilir et al.(2006)]{Bilir06} 
Bilir, S., G{\"u}ver, T., Aslan, M., 2006, Astron. Nachr., 327, 693

\bibitem[Bilir et al.(2010)]{Bilir10}
Bilir, S., G{\"u}ver, T., Khamitov, I., Ak, T., Ak, S., Co{\c s}kuno{\u g}lu, K.~B., Paunzen, E., Yaz, E., 2010, \apss, 326, 139 
 
\bibitem[Bilir et al.(2016)]{Bilir16}
Bilir, S., Bostanc{\i}, Z. F., Yontan, T., et al., 2016, Adv. Space Res., 58, 1900

\bibitem[Bisht et al.(2020)]{Bisht20} 
Bisht, D., Zhu, Q., Yadav, R.~K.~S., Durgapal, A., Rangwal, G., 2020, \mnras, 494, 607 

\bibitem[Bossini et al.(2019)]{Bossini19}
Bossini, D., Vallenari, A., Bragaglia, A., et al., 2019, \aap, 623, A108. 

\bibitem[Bostanc{\i} et al.(2015)]{Bostanci15}
Bostanc{\i}, Z.~F., Ak, T., Yontan, T., et al., 2015, \mnras, 453, 1095 

\bibitem[Bostanc{\i} et al.(2018)]{Bostanci18}
Bostanc{\i}, Z.~F., Yontan, T., Bilir, S., et al., 2018, \apss, 363, 143 

\bibitem[Bovy \& Tremaine(2012)]{Bovy12}
Bovy, J., Tremaine, S., 2012, \apj, 756, 89

\bibitem[Bovy(2015)]{Bovy15}
Bovy, J., 2015, \apjs, 216, 29

\bibitem[Bressan et al.(2012)]{Bressan12}
Bressan, A., Marigo, P., Girardi, L., Salasnich, B., Dal Cero, C., Rubele, S., Nanni, A., 2012, \mnras, 427, 127

\bibitem[Cannon \& Lloyd(1969)]{Cannon69}
Cannon, R.~D., Lloyd, C., 1969, \mnras, 144, 449

\bibitem[Cantat-Gaudin et al.(2018)]{Cantat-Gaudin18}
Cantat-Gaudin, T., Jordi, C., Vallenari, A., Bragaglia, A., et al., 2018, \aap, 618, 93

\bibitem[Cantat-Gaudin \& Anders(2020)]{Cantat-Gaudin_Anders20}
Cantat-Gaudin, T., Anders, F., 2020, \aap, 633A, 99 

\bibitem[Cantat-Gaudin et al.(2020)]{Cantat-Gaudin20}
Cantat-Gaudin, T., Anders, F., Castro-Ginard, A., Jordi, C., et al., 2020, \aap, 640, A1

\bibitem[Canterna et al.(1986)]{Canterna86}
Canterna, R., Geisler, D., Harris, H.~C., Olszewski, E., Schommer, R., 1986, \aj, 92, 79

\bibitem[Carraro \& Chiosi(1994)]{Carraro94}
Carraro, G., Chiosi, C., 1994, \aap, 288, 751

\bibitem[Carraro et al.(2017)]{Carraro17}
Carraro, G., Sales Silva, J. V., Moni Bidin, C., Vazquez, R. A., 2017, \apj, 153, 20

\bibitem[Carrera et al.(2019)]{Carrera19}
Carrera, R., Bragaglia, A., Cantat-Gaudin, T., et al., 2019, \aap, 623, A80 

\bibitem[Casamiquela et al.(2017)]{Casamiquela17}
Casamiquela, L., Carrera, R., Blanco-Cuaresma, et al., 2017, \mnras, 470, 4363

\bibitem[Casamiquela et al.(2019)]{Casamiquela19}
Casamiquela, L., Blanco-Cuaresma, S., Carrera, R., et al., 2019, \mnras, 490, 1821 

\bibitem[Castro-Ginard et al.(2020)]{Castro-Ginard20}
Castro-Ginard, A., Jordi, C., Luri, X., et al., 2020, \aap, 635, A45

\bibitem[Chincarini(1963)]{Chincarini63}
Chincarini, G., 1963, CoAsi, 138, 19

\bibitem[Chen et al.(2000)]{Chen00}
Chen, Y.~Q., Nissen, P. E., Zhao, G., Zhang, H.~W., Benoni, T., 2000, Astron. Astrophys. Suppl. Ser., 141, 491

\bibitem[Conrad et al.(2017)]{Conrad17}
Conrad, C., Scholz, R.-D., Kharchenko, N.~V., et al., 2017, \aap, 600, A106

\bibitem[Co{\c{s}}kuno{\v{g}}lu et al.(2011)]{Coskunoglu11}
Co{\c{s}}kuno{\v{g}}lu, B., Ak S., Bilir, S., et al., 2011, \mnras, 412, 1237

\bibitem[Cuffey(1944)]{Cuffey44}
Cuffey, J., 1944, \aj, 51, 65

\bibitem[Dias et al.(2002)]{Dias02}
Dias, W. S., Alessi, B. S., Moitinho, A., Lepine, J. R. D., 2002, \aap, 389, 871

\bibitem[Dias et al.(2014)]{Dias14} 
Dias, W.~S., Monteiro, H., Caetano, T.~C., L{\'e}pine, J.~R.~D, Assafin, M., Oliveira, A.~F., 2014, \aap, 564, A79

\bibitem[Dias et al.(2021)]{Dias21}
Dias, W.~S., Monteiro, H., Moitinho, A., et al., 2021, \mnras, 504, 356

\bibitem[Donor et al.(2020)]{Donor20}
Donor, J., Frinchaboy, P.~M., Cunha, K., et al., 2020, \aj, 159, 199. 

\bibitem[Eker et al.(2018)]{Eker18}
Eker, Z., Bak\i \c s, V., Bilir, S., et al., 2018, \mnras, 479, 5491

\bibitem[Eker et al.(2020)]{Eker20}
Eker, Z., Soydugan, F., Bilir, S., et al., 2020, \mnras, 496, 3887 

\bibitem[Fang(1970)]{Fang70}
Fang, C., 1970, \aap, 4, 75

\bibitem[Friel et al.(2002)]{Friel02} 
Friel, E.~D., Janes, K.~A., Tavarez, M., et al., 2002, \aj, 124, 2693

\bibitem[Gaia collaboration et al.(2018)]{Gaia18}
Gaia Collaboration, Brown, A. G. A., Vallenari, A., Prusti, T., et al., 2018, \aap, 616, 22 

\bibitem[Gaia collaboration et al.(2021)]{Gaia21}
Gaia Collaboration, Brown, A. G. A., Vallenari, A., Prusti, T., et al., 2021, \aap, 694, 1

\bibitem[Garcia, Claria \& Levato(1988)]{Garcia88} 
Garcia, B., Claria, J.~J., Levato, H., 1988, \apss, 143, 317.

\bibitem[Geisler(1988)]{Geisler88}
Geisler, D., 1988, \pasp, 100, 338

\bibitem[Geisler, Claria \& Minniti(1991)]{Geisler91}
Geisler, D., Claria, J.~J., Minniti, D., 1991, \aj, 102, 1836

\bibitem[Gilmore et al.(2012)]{Gilmore12}
Gilmore, G., Randich, S., Asplund, M., et al., 2012, Msngr, 147, 25

\bibitem[Glushkova \& Rastorguev(1991)]{Glushkova91}
Glushkova, E.~V., Rastorguev, A.~S., 1991, SvAL, 17, 64

\bibitem[Groenewegen(2021)]{Groenewegen21}
Groenewegen, M.~A.~T., 2021, \aap, 654, A20

\bibitem[Hao et al.(2021)]{Hao21}
Hao, C.~J., Xu, Y., Hou, L.~G., et al., 2021, \aap, 652, A102

\bibitem[Hoag \& Applequist(1965)]{Hoag65}
Hoag, A.~A., Applequist, N.~L., 1965, \apjs, 12, 215

\bibitem[Jacobson, Friel \& Pilachowski(2007)]{Jacobson07}
Jacobson, H.~R., Friel, E.~D., Pilachowski, C.~A., 2007, \aj, 134, 1216

\bibitem[Jacobson \& Friel(2013)]{Jacobson13}
Jacobson, H.~R., Friel, E.~D., 2013, \aj, 145, 107

\bibitem[Jadhav \& Subramaniam (2021)]{Jadhav21} 
Jadhav, V.~V., Subramaniam, A., 2021, \mnras, 507, 1699 

\bibitem[Janes \& Hoq(2011)]{Janes11}
Janes, K., Hoq, S., 2011, \aj, 141, 92 

\bibitem[Joshi et al.(2016)]{Joshi16}
Joshi, Y.~C., Dambis, A.~K., Pandey, A.~K., Joshi, S., 2016, \aap, 593, A116

\bibitem[Karaali et al.(2003a)]{Karaali03a}
Karaali, S., Bilir, S., Karata{\c{s},} Y., Ak, S.~G., 2003a, \pasa, 20, 165

\bibitem[Karaali et al.(2003b)]{Karaali03b}
Karaali, S., Ak, S.~G., Bilir, S., Karata{\c{s}}, Y., Gilmore, G., 2003b, \mnras, 343, 1013.

\bibitem[Karaali et al.(2011)]{Karaali11}
Karaali, S., Bilir, S., Ak, S., Yaz, E., Co\c skuno\u glu, B., 2011, \pasa, 28, 95

\bibitem[Kharchenko et al.(2005)]{Kharchenko05}
Kharchenko, N.~V., Piskunov, A.~E., R{\"o}ser, S., Schilbach, E., Scholz, R.-D., 2005, \aap, 438, 1163. 

\bibitem[Kharchenko et al.(2012)]{Kharchenko12}
Kharchenko, N. V., Piskunov, A. E., Schilbach, E., Roeser, S., Scholz, R. -D., 2012, \aap, 543, 156 

\bibitem[Kharchenko et al.(2013)]{Kharchenko13} 
Kharchenko, N.~V., Piskunov, A.~E., Schilbach, E., R{\"o}ser, S., Scholz, R.-D., 2013, \aap, 558, A53

\bibitem[Kharchenko et al.(2016)]{Kharchenko16}
Kharchenko, N.~V., Piskunov, A.~E., Schilbach, E., R{\"o}ser, S., Scholz, R.-D., 2016, \aap, 585, A101

\bibitem[King(1962)]{King62}
King, I., 1962, \aj, 67, 471

\bibitem[K\"ustner(1923)]{Kustner23}
Kustner, F., 1923, VeBon, 18, 1

\bibitem[Lada \& Lada(2003)]{Lada03} 
Lada, C.~J., Lada, E.~A., 2003, ARA\&A, 41, 57

\bibitem[Landolt(2009)]{Landolt09}
Landolt, A. U., 2009, \aj, 137, 4186

\bibitem[Lang(2010)]{Lang10}
Lang, D., Hogg, D.~W., Mierle, K., Blanton, M., Roweis, S., 2010, AJ, 139, 1782 

\bibitem[Larsson-Leander(1957)]{Larsson57} 
Larsson-Leander, G., 1957, StoAn, 20, 1

\bibitem[Leggett(1992)]{Leggett92}
Leggett, S.~K., 1992, \apjs, 82, 351

\bibitem[Lindoff(1968)]{Lindoff68}
Lindoff, U., 1968, ArA, 5, 1

\bibitem[Liu \& Pang(2019)]{Liu19}
Liu, L., Pang, X., 2019, \apjs, 245, 32

\bibitem[Maciejewski \& Niedzielski(2007)]{Maciejewski07}
Maciejewski, G., Niedzielski, A., 2007, \aap, 467, 1065  
 
\bibitem[Mermilliod, Huestamendia \& del Rio(1994)]{Mermilliod94}
Mermilliod, J.-C., Huestamendia, G., del Rio, G., 1994, Astron. Astrophys. Suppl. Ser., 106, 419

\bibitem[Milone(1994)]{Milone94} 
Milone, A.~A.~E., 1994, \pasp, 106, 1085

\bibitem[Miyamoto \& Nagai(1975)] {Miyamoto75}
Miyamoto, M., Nagai, R., 1975, \pasj, 27, 533

\bibitem[Moraux(2016)]{Moraux16}
Moraux, E., 2016, EAS, 80-81, 73

\bibitem[Navarro et al.(1996)] {Navarro96}
Navarro, J.~F., Frenk, C.~S., White, S.~D.~M., 1996, \apj, 462, 563

\bibitem[Netopil, Paunzen \& Carraro(2015)]{Netopil15}
Netopil, M., Paunzen, E., Carraro, G., 2015, \aap, 582, A19

\bibitem[Rain et al.(2021)]{Rain21} 
Rain, M.~J., Carraro, G., Ahumada, J.~A., Villanova, S., Boffin, H., Monaco, L., 2021, AJ, 161, 37

\bibitem[Reddy, Lambert \& Giridhar(2016)]{Reddy16}
Reddy, A.~B.~S., Lambert, D.~L., Giridhar, S., 2016, \mnras, 463, 4366

\bibitem[Robb \& Cardinal(1998)]{Robb98}
Robb, R.~M., Cardinal, R.~D., 1998, IBVS, 4634, 1

\bibitem[Rosvick \& Balam(2002)]{Rosvick02}
Rosvick, J.~M., Balam, D., 2002, \aj, 124, 2093

\bibitem[R{\"o}ser, Demleitner \& Schilbach(2010)]{Roeser10}
R{\"o}ser, S., Demleitner, M., Schilbach, E., 2010, \aj, 139, 2440

\bibitem[Salpeter(1955)]{Salpeter55}
Salpeter, E.~E., 1955, \apj, 121, 161

\bibitem[Sampedro et al.(2017)]{Sampedro17} 
Sampedro, L., Dias, W.~S., Alfaro, E.~J., Monteiro, H., Molino, A., 2017, \mnras, 470, 3937

\bibitem[Science Software Branch at STScI(2012)]{Science12}
Science Software Branch at STScI, 2012, PyRAF: Python alternative for IRAF,
Astrophysics Source Code Library, ascl:1207.011

\bibitem[Sim et al.(2019)]{Sim19} 
Sim, G., Lee, S.~H., Ann, H.~B., Kim, S., 2019, J. Korean Astron. Soc., 52, 145

\bibitem[Soubiran et al.(2018)]{Soubiran18}
Soubiran, C., Cantat-Gaudin, T., Romero-G{\'o}mez, M., et al., 2018, \aap, 619, A155

\bibitem[Sun et al.(2021)]{Sun21} 
Sun, M., Jiang, B., Yuan, H., Li, J., 2021, ApJS, 254, 38

\bibitem[Spitzer \& Hart(1971)]{Spitzer71}
Spitzer, L., Hart, M. H., 1971, \apj, 164, 399

\bibitem[Stetson(1980)]{Stetson80} 
Stetson, P.~B., 1980, \aj, 85, 387

\bibitem[Sung et al.(2013)]{Sung13}
Sung, H., Lim, B., Bessell, M. S., Kim, J. S., Hur, H., Chun, M., Park, B., 2013, J. Korean Astron. Soc., 46, 103

\bibitem[Tarricq et al.(2021)]{Tarricq21}
Tarricq, Y., Soubiran, C., Casamiquela, L.,  et al., 2021, \aap, 647, A19 

\bibitem[Thogersen \& Fallon(1993)]{Thogersen93} 
Thogersen, E.~N., Friel, E.~D., Fallon, B.~V., 1993, \pasp, 105, 1253

\bibitem[Tody(1986)]{Tody86}
Tody, D., 1986, SPIE, 627, 733

\bibitem[Tody(1993)]{Tody93}
Tody, D., 1993, ASPC, 52, 173

\bibitem[Trumpler(1930)]{Trumpler30}
Trumpler, R.~J., 1930, LicOB, 420, 154

\bibitem[Warren \& Cole(2009)]{Warren09}
Warren, S.~R., Cole, A.~A., 2009, \mnras, 393, 272

\bibitem[Yadav, Sariya \& Sagar(2013)]{Yadav13} 
Yadav, R.~K.~S., Sariya, D.~P., Sagar, R., 2013, \mnras, 430, 3350

\bibitem[Yontan et al.(2015)]{Yontan15}
Yontan, T., Bilir, S., Bostanc\i, Z. F., et al., 2015, \apss, 355, 267 

\bibitem[Yontan et al.(2019)]{Yontan19}
Yontan, T., Bilir, S., Bostanc\i, Z. F., et al., 2019, \apss, 364, 20

\bibitem[Yontan et al.(2021)]{Yontan21}
Yontan, T., Bilir, S., Ak, T., et al., 2021, Astron. Nachr., 342, 538

\bibitem[Zhai et al.(2017)]{Zhai17}
Zhai, M., Abt, H., Zhao, G., Li, C., 2017, \aj, 153, 57.

\bibitem[Zhao \& He(1990)]{Zhao90} 
Zhao, J.~L., He, Y.~P., 1990, \aap, 237, 54
\end{thebibliography}
\end{document}